\documentclass[%
 reprint,
superscriptaddress,
 amsmath,amssymb,
 aps,
long bibliography,
prl
]{revtex4-2}

\usepackage[utf8]{inputenc}

\usepackage[dvipsnames,svgnames,table]{xcolor}
\usepackage{verbatim}
\usepackage{graphicx}
\usepackage{bm}
\usepackage{hyperref}
\usepackage{mathtools}
\hypersetup{nolinks=true}
\usepackage{siunitx}
\usepackage[capitalise]{cleveref}
\usepackage[super]{nth}

\usepackage{tikz}

\newcommand{\vect}[1]{\ensuremath{{\bm{#1}}}}
\newcommand{\mat}[1]{\mathbf{#1}}
\newcommand{\grad}{\vect{\nabla}}
\newcommand{\gradperp}{\vect{\nabla}_\perp}
\newcommand{\xperp}{\vect{x}_\perp}

\newcommand{\partiald}[2]{\ensuremath{\frac{\partial#1}{\partial#2}}}

\newcommand{\adim}[1]{{#1}}
\newcommand{\ddim}[1]{\tilde{#1}}

\newcommand{\OO}[1]{\mathcal{O}({#1})}

\newcommand{\revision}[1]{\textcolor{black}{{#1}}}
\newcommand{\rrevision}[1]{\textcolor{black}{{#1}}}

\graphicspath{{./FIGURES}}

\begin{document}

\title{Hovering of an actively driven fluid-lubricated foil}

\author{Stéphane Poulain}
\affiliation{%
	Department of Mathematics, University of Oslo, 0851 Oslo, Norway
}
\author{Timo Koch}
\affiliation{%
	Department of Mathematics, University of Oslo, 0851 Oslo, Norway
}
\author{L. Mahadevan}
\email{lmahadev@g.harvard.edu}
\affiliation{%
	School of Engineering and Applied Sciences, Harvard University, Cambridge, Massachusetts 02138, USA
}
\affiliation{%
	Department of Physics and Department of Organismic and Evolutionary Biology, Harvard University, Cambridge, Massachusetts 02138, USA
}
\author{Andreas Carlson}
\email{acarlson@math.uio.no}
\affiliation{%
	Department of Mathematics, University of Oslo, 0851 Oslo, Norway
}


\begin{abstract}
Inspired by recent experimental observations of a harmonically excited elastic foil hovering near a wall while supporting substantial weight, we develop a theoretical framework that describes the underlying physical effects.
Using elastohydrodynamic lubrication theory, we quantify how the dynamic deformation of the soft foil couples to the viscous fluid flow in the intervening gap. Our analysis shows that the soft foil rectifies the reversible forcing, breaking time-reversal symmetry; the spatial distribution of the forcing determines whether the sheet is attracted to or repelled from the wall. A simple scaling law predicts the time-averaged equilibrium hovering height and the maximum weight the sheet can sustain before detaching. Numerical simulations of the governing equation corroborate our theoretical predictions, are in qualitative agreement with experiments, and might explain the behavior of organisms while providing design principles for soft robotics.
\end{abstract}

\maketitle

Hovering near surfaces has evolved in animals across diverse environments: insects hover above water and plants, birds above land and water, and fish in benthic environments and near other animals~\cite{Vogel}.
This phenomenon has long inspired visionary science fiction writers and engineers alike, and hovering-based technology is now employed in various applications~\cite{Brandt1989}.
Although hovering flight is primarily associated with large-scale, high Reynolds number flows, it is not restricted to this regime.
In viscous fluids, soft objects moving along a wall create lift forces through elastohydrodynamic coupling~\cite{Skotheim2004}, a mechanism with a wide range of applications in biology, microfluidics, and nanoscience \cite{Leroy2012,SaintYves2005,Essink2021,Fares2024, Bureau2023,Rallabandi2024}.
Also, hydrodynamic interactions coupled with unsteady elastic deformations of a foil enable its levitation above surfaces, even in the absence of inertia \cite{Argentina2007, Jafferis2011}.

Recent experiments reveal a striking manifestation of related ideas in contactless robotic manipulation, whereby an actively driven elastic foil lifts a heavy load while hovering just below a rigid surface, akin to a contactless suction cup ~\cite{Weston2021}.
Current explanations for this hovering are based on compressible and inertial effects in the surrounding fluid \cite{Ramanarayanan2022}, but overlook the potentially crucial effects generated by elastic deformations of the foil.
Here, we combine scaling estimates, asymptotic models, and numerical simulations of the governing equations to show that accounting for the underlying viscous elastohydrodynamics is critical in explaining the hovering of actively driven foils.
We reveal how a periodic drive can lead to an aperiodic response, and predict the average height of hovering and the maximum load that can be sustained, consistent with experimental observations.

\begin{figure}
    \includegraphics[width=\columnwidth]{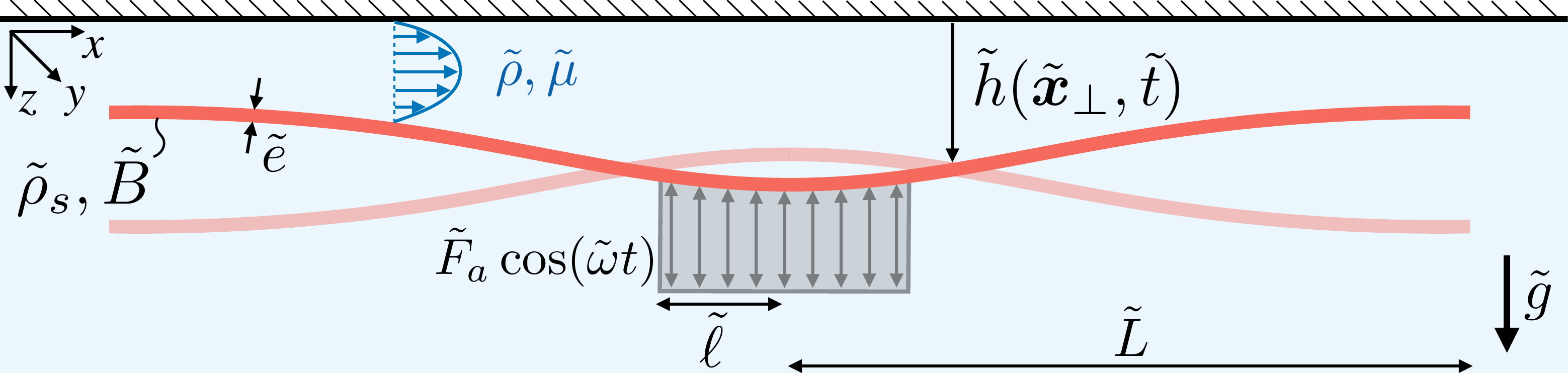}\vspace{2mm}
    \caption{Schematic of an elastic sheet of bending rigidity $\tilde B$, radius $\tilde L$, thickness $\tilde e$ and density $\tilde \rho_s$ immersed in a fluid of density $\tilde \rho$ and viscosity $\tilde \mu$. The sheet bends in response to a harmonic normal force $\tilde F_a \cos(\tilde \omega \tilde t)$ distributed over an area of radius $\tilde \ell$, which drives a flow in the thin gap.
}
  \label{fig:schematics}
\end{figure}

\paragraph{Setup.}
To describe hovering near a wall, we consider an elastic sheet of radius $\ddim L$ subjected to a normal harmonic force of amplitude $\tilde F_a$ and angular frequency $\tilde \omega$.
The associated load $\tilde p_a \cos(\tilde \omega \tilde t)$ is uniformly distributed over a disk of radius $\tilde \ell$, see \cref{fig:schematics}.
The sheet has a density $\ddim \rho_s$, Young's modulus $\ddim E$, Poisson's ratio $\nu$,
and thickness $\ddim e$; its bending modulus is $\ddim B = \ddim E \ddim e^3/12(1-\nu^2)$ \revision{and its weight is $\tilde W$}.
The gravitational acceleration $\ddim g$ is normal to the wall.
The surrounding incompressible fluid has a dynamic viscosity $\tilde \mu$ and density $\tilde \rho$.
Experiments \cite{Weston2021} have been performed  in air ($\tilde \mu \simeq 2\times 10^{-5} \unit{\pascal\second}$, $\tilde \rho \simeq 1.2\unit{\kilo\gram\per\meter\cubed} $) using thin  sheets ($\tilde e\simeq 300\unit{\micro\meter}$, $\tilde L\simeq10\unit{\centi\meter}$) made of plastic ($\tilde E\simeq 3\unit{\giga\pascal}$, $\nu \simeq 0.3$, $\tilde \rho_s \simeq 1400\unit{\kilo\gram\per\meter\cubed}$).
An eccentric mass motor ($\tilde \ell \simeq 1\unit{\centi\meter}$), i.e.,  a mass $\tilde m \simeq 0.4\unit{\gram}$ rotating with $\tilde \omega\simeq 2\pi \times 200 \unit{\hertz}$ and with a gyration radius $\tilde r \simeq 1\unit{\milli\meter}$, provides a force $\tilde F_a=\tilde m \tilde r \tilde \omega^2 \simeq 1\unit{\newton}$.

\paragraph{\revision{Scaling analysis.}}
\revision{We introduce the characteristic width $\ddim L$ and height $\ddim H$ of the system. 
\rrevision{We note that $\tilde H$ serves as a scale for the gap thickness, vertical oscillation amplitude, and sheet deformation, all of which are a priori unknown.}
We construct characteristic velocities  $\ddim \omega \ddim H$  and $\ddim \omega \ddim L$ in the vertical and horizontal directions, respectively.
We assume that the fluid-filled gap between the sheet and the wall is narrow (\cref{fig:schematics}), $\tilde H/\tilde L \ll 1$, and that inertial effects are negligible. This, combined with a small film Reynolds number based on characteristic length $\tilde H$ and velocity $\ddim \omega \ddim H$, ${\rm Re}=\tilde \rho \tilde \omega \tilde H^2/\tilde \mu \ll 1$, justifies the use of lubrication theory  \cite{Batchelor}.}
\revision{On dimensional grounds,} the viscous pressure in the gap scales as $\revision{\tilde p_v=}\ddim \mu \ddim \omega \tilde L^2/\tilde H^2$ \cite{Batchelor}, and the sheet's bending pressure as $\revision{\tilde p_b=}\ddim B \ddim H/\ddim L^4$.
Two dimensionless quantities characterize the periodic actuation: $\Gamma=\tilde p_a/\tilde p_v$ the ratio of active and viscous stresses, and $\gamma=\tilde p_a/\tilde p_b$ the ratio of active and bending stresses.

\revision{Experiments show that the actively driven sheet can sustain a weight $\tilde W$ (with weight per unit area $\tilde p_w=\tilde W/\tilde L^2$) at a time-averaged equilibrium height away from the wall \cite{Weston2021}.}
We expect the maximum supported weight to increase with active stress and to decrease with bending pressure \rrevision{(in the limit of large bending stresses)}, since either a passive or rigid sheet cannot support any weight.
As  $\tilde W$ may only depend on even powers of $\tilde p_a$ (reversing the sign of $\ddim p_a$ is equivalent to a phase shift, which cannot affect the long-term dynamics),
\revision{this leads us to define a dimensionless weight $\mathcal W = \tilde p_w \tilde p_b /\tilde p_a^2$.}
In what follows, we verify these \revision{heuristic} scaling arguments and demonstrate, using asymptotic analysis and numerical simulations, how active soft sheets are attracted to or repelled from surfaces, and how viscous elastohydrodynamics enable stable hovering at heights scaling as $\tilde H_{\rm bv} \sim \ddim L^2 (\ddim \mu \ddim \omega/\ddim B)^{1/3}$, the characteristic height defined by balancing bending and viscous stresses (\rrevision{$\tilde p_v=\tilde p_b$, $\Gamma=\gamma$}).
\rrevision{In particular, we find that the equilibrium hovering height $\tilde h_{\rm eq} $ is such that $\tilde h_{\rm eq}/\tilde H_{\rm bv}$ is a universal function of the dimensionless weight $\mathcal W$ in the limit of weak forcing ($\gamma \lesssim 1$).}

\paragraph{Governing equations.}
We use lubrication theory \cite{Batchelor} to describe the fluid flow.
Balancing the horizontal pressure gradient with the transverse viscous stresses yields an evolution equation for $\ddim h(\tilde{\vect x}_\perp,\ddim t)$, the distance
between the sheet and the wall at the position $\tilde{\vect x}_\perp=(\tilde x, \tilde y)$ (\cref{fig:schematics}):
\begin{align}
    \partiald{ h}{t} -  \gradperp\cdot \left(\frac{h^3\gradperp p}{12} \right) = 0,
    \label{eq:NS}
\end{align}
with $\vect q=-h^3\gradperp p\rrevision{/12}$ the horizontal volumetric flux.
The governing equations and results are presented in dimensionless units (written without tilde throughout the Letter), with 
$({\vect x}_\perp,\ell)=(\ddim{{\vect x}}_\perp,\ddim \ell)/\ddim L$, $\adim t = \ddim t\ddim\omega$, $\adim h=\ddim h/\ddim H$, $\adim p = \ddim p/\ddim p_v$.
The gauge pressure $\tilde p$ is measured relative to the atmospheric pressure, and
$\gradperp=(\partial/\partial x,~\partial/\partial y)$ is the horizontal gradient.
\rrevision{We neglect inertial effects and consider only bending deformations of the sheet. The assumptions of small deformations and lubrication theory make tension in the sheet negligible, as further described in the Supplemental Material \cite{supp}}. The normal stress balance then follows Kirchhoff-Love theory~\cite{Timoshenko1959,Landau1986}:
\begin{align}
\begin{split}
	&-p = \gradperp \cdot \left(\grad_\perp \cdot \mat M\right)  +  f_a\left(\xperp,t\right) + \gamma \Gamma \mathcal W , \label{eq:forcebalance}\\
    & \mat M = -{\frac\Gamma\gamma}\left[(1-\nu)\boldsymbol \kappa + \nu{\rm tr}\left(\boldsymbol \kappa \right) \mathbf I\right], 
\end{split}
\end{align}
with  $\mat M$ the matrix of bending moments and $\boldsymbol \kappa$ the Hessian of $h$, the sheet's local curvature.
\rrevision{We assume that} the active stress $f_a$ is distributed uniformly around the center of the sheet (\cref{fig:schematics}): $f_a(\xperp,t)=\Gamma\cos(t)/\ell$ if $\lvert \xperp \rvert < \ell$, $=0$ otherwise.
We expect qualitatively the same behavior for a one-dimensional (1D) and two-dimensional (2D) system and focus, for simplicity, the subsequent analyses on a 1D sheet: $\xperp \rightarrow x$, $\gradperp \rightarrow \partial/\partial x$, $\partial/\partial y = 0$. 
As boundary conditions, we use the fact that the sheet's edges are stress-free, torque-free, and at atmospheric pressure: $p=\partial^2 h/\partial x^2=\partial^3 h/\partial x^3=0$ for $x=\pm 1$. 

\begin{figure}
    \includegraphics[width=0.9\linewidth]{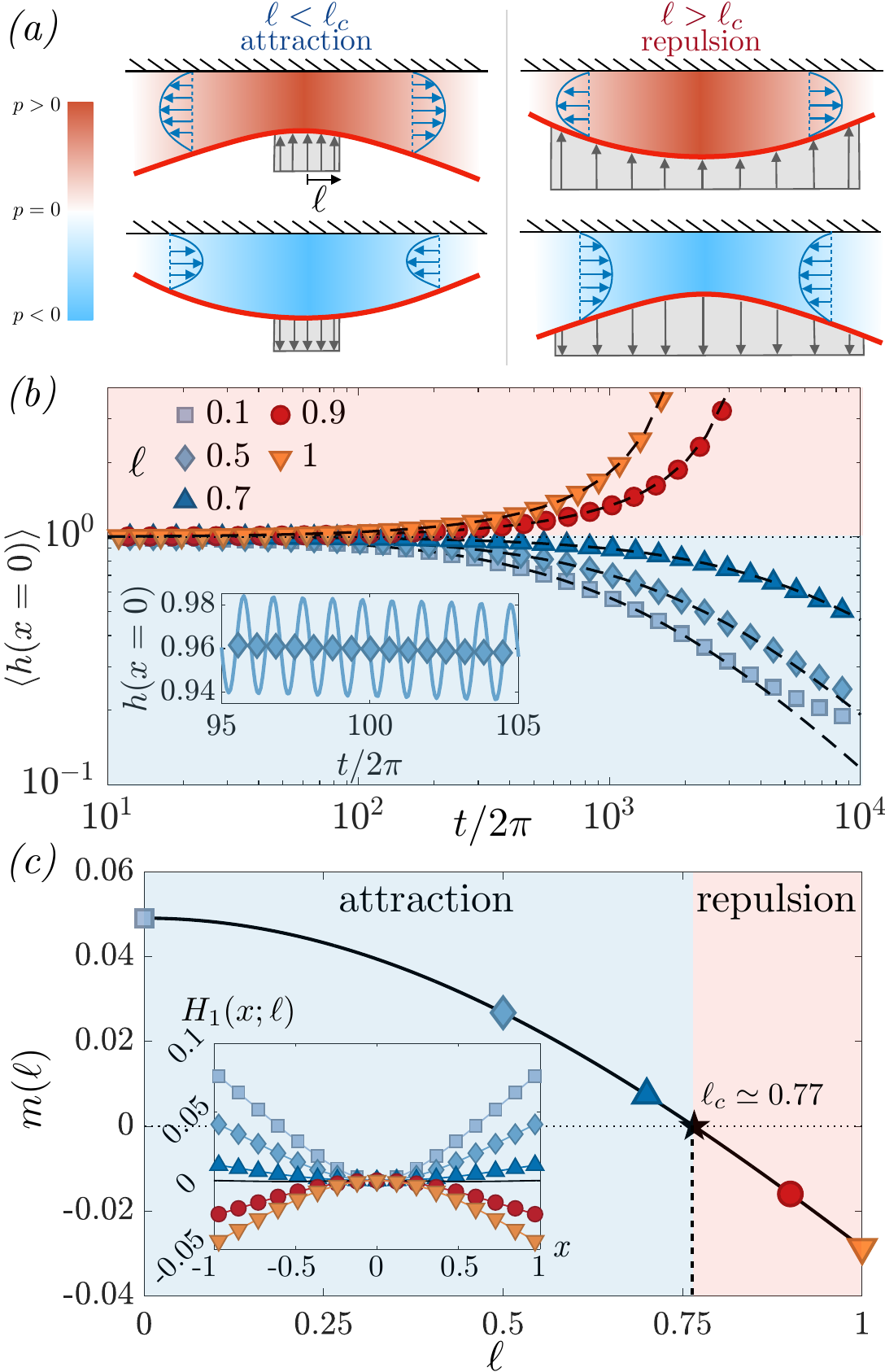}
    \caption{%
    $(a)$ Schematic of the sheet's deformation  based on $\ell $.
    See also Supplementary Movies S1 and S2 \cite{supp}.
    $(b)$ Numerical solutions of $\eqref{eq:NS}$ and $\eqref{eq:forcebalance}$ for the time-evolution of the averaged height (symbols) for $\gamma=0.1,~\mathcal W=0$. Dashed lines are the asymptotic result \eqref{eq:soft}.
    The inset shows a zoom for $\ell=0.5$ that highlights the slow average compared to the fast oscillations (solid line).
    $(c)$ The function $m(\ell)$ appears in \eqref{eq:soft} and determines whether the sheet is attracted  ($m(\ell)>0$) or repelled $(m(\ell)<0)$ from the wall. 
    The shape of the sheet at $\OO{\gamma}$  is $H_{1}(x;\ell)\cos(t)$, characterized in the inset.
    }
\label{fig:soft}
\end{figure}

\paragraph{\revision{Large distances ($\tilde h \gg \tilde{H}_{\rm bv}$)}.}
We first consider a weightless foil ($\mathcal W=0$), which allows us to study the effect of the sheet's softness in isolation.
We characterize the magnitude of the sheet's deformation with  $\gamma$, the ratio of active and bending pressures as defined above.
\rrevision{Here we define the height scale $\tilde H$ as the initial height of the sheet, $\tilde H=\tilde h(\tilde t=0)$, and such that $\tilde H \gg \tilde H_{\rm bv}$ ($\gamma \ll \Gamma$, $\tilde p_b \gg \tilde p_v$).}
In the limit $\gamma \to 0$\revision{, $\Gamma$ finite},  the sheet is flat and rigid, and the film height is only a function of time: 
combining \eqref{eq:NS} with \eqref{eq:forcebalance} \rrevision{integrated in space} yields $p(x,t)=3\Gamma  \cos(t) \left(1-x^2\right)/2$ and $h(t)=\left(1+\Gamma  \sin (t)/2\right)^{-1/2}$.
The time-averaged height $\langle h \rangle (t) = \int_t^{t+2\pi}h(t)~{\rm d}t/2\pi$ is constant and the dynamics is time-reversible.
To predict the behavior when the sheet deforms, $\gamma>0$, we integrate \eqref{eq:NS} over the length of the sheet and average in time to find $\langle \partial h /\partial t\rangle = -\langle q_e\rangle$, with $\langle q_e \rangle=  -\langle h^3 \partial p/\partial x \rangle_{x=1}\rrevision{/12}$ the time-averaged flux at the edge of the sheet.
We expect that the sheet's elastic response at leading order is in phase with the forcing:  $h(x,t) \simeq h_0(t)+\gamma \cos(t) H_{1}(x;\ell)$, with $H_{1}$  describing the sheet's deformation, which depends on the relative extent $\ell$ of the forcing.
By using the pressure distribution obtained for the rigid sheet in the evaluation of the flux, we then find at leading order
\begin{align}
    \partiald{ h_0 }{t} \sim -{\gamma \Gamma} H_{1}(1;\ell) h_0 ^2,
    \label{eq:avg}
\end{align}
\rrevision{with $h_0 \simeq \langle h_0 \rangle$ for a slowly-varying time-evolution of the height}.
The edge deformation $H_{1}(1;\ell)$ is determined by the shape the sheet takes when it is subjected on one side to the parabolic fluid pressure across its entire length and on the other side to the active rectangular forcing of length $2\ell$ (\cref{fig:soft}a).
As $\ell \rightarrow 0$, the sheet's edges are influenced only by fluid pressure and bend accordingly: $H_1(1;\ell)>0$. Conversely, as $\ell \rightarrow 1$, the active forcing dominates the edges, and the sheet bends in the opposite direction.
This indicates a critical length $\ell_c$ at which $H_1(1;\ell)$ changes sign,
with the sheet attracted to the wall for $\ell < \ell_c$ 
and repelled otherwise.
A naïve estimate suggests that this transition occurs when the fluid pressure at the center, $3 \Gamma \cos(t)/2$, and the active stress, $\Gamma \cos(t)/\ell$, are similar, i.e. $\ell \simeq 2/3$; a more careful calculation below confirms that this is a reasonable estimate. 

To verify and go beyond these scaling predictions, we solve \eqref{eq:NS} and \eqref{eq:forcebalance} numerically \cite{Koch2021,supp}. \Cref{fig:soft}$(a)$ and the supplementary movies S1-S3 \cite{supp} illustrate the coupling between the sheet's deformation and pressure distribution that leads to attraction or repulsion. \Cref{fig:soft}$(b)$ shows that the sheet's averaged motion, characterized by $\langle h(0,t)\rangle$, is slow compared to the periodic forcing.
This observation motivates a two-timescale analysis of the governing equations.
We assume \rrevision{$\gamma\ll 1$, $\Gamma=\OO{\gamma^0}$, $\mathcal W=\OO{\gamma^0}$}, and that the dynamics depend both on the time $t$ associated with the active forcing and a slow time $\tau = \gamma t$ describing the averaged evolution: $h(x,t)$ becomes $h(x,t,\tau)$. This allows us to treat the sheet's deformation as a small perturbation to the response of a forced rigid sheet.
\rrevision{We note that we also require $h=\OO{\gamma^0}$, i.e., $\tilde h  \sim \tilde H \gg \tilde H_{\rm bv}$.}
We then expand $h$ in powers of $\gamma$,
$h(x,t,\tau)=h_0(t,\tau)+\gamma h_1(x,t,\tau)+ \gamma^2 h_2(x,t,\tau)$, and the time derivative as $\partial/\partial t \rightarrow \partial/\partial t + \gamma \partial/\partial \tau$.
At $\OO{\gamma^2}$, we find for the sheet's oscillations \cite{supp}:
\begin{subequations}
\begin{equation}
\begin{split}
    h_0(t,\tau)&=\left(f(\tau)+\frac\Gamma2\sin(t)\right)^{-1/2}, \\
    \frac1\Gamma \frac{{\rm d}f}{{\rm d} \tau }(\tau)&=m(\ell)f(\tau)^{1/2}\revision{-\frac{\mathcal W}{2}},
\end{split}
\label{eq:soft}
\end{equation}
and for its deformation:
\begin{equation}
\begin{split}
    h_1(x,t,\tau)&=\cos(t)H_1(x;\ell)+h_1'(t,\tau), \\
    h_2(x,t,\tau)&=\revision{\mathcal W H_1(x;1)}+\frac{\sin(t)}{\Gamma  h_0^3(t,\tau)}H_{2}(x;\ell)+\\
    &\qquad \qquad \qquad \qquad \quad +\frac{\cos^2(t)}{h_0(t,\tau)}H_{2}^\star(x;\ell).
\end{split}
\label{eq:soft2}
\end{equation}
\label{eq:softall}
\end{subequations}
The analytical expression of $m(\ell)$, $H_{1}(x;\ell)$, $H_{2}(x;\ell)$ and $H_{2}^\star(x;\ell)$ are given in \cite{supp}.
The function $h_1'$ has zero mean and does not contribute to the time-averaged dynamics.

\revision{We first consider \eqref{eq:softall} without gravity, $\mathcal W=0$. $f(\tau)$ characterizes the slow evolution of the system that can be integrated and yields the time-averaged sheet's center height:} $\langle h \rangle (0,t) \simeq (1+0.5\gamma\Gamma m(\ell)t)^{-1}$, with $h(x,0)=\rrevision{1}$.
This agrees with our predicted scaling \eqref{eq:avg} and with the numerical simulations, as shown in \cref{fig:soft}($b$).
The nature of the sheet's motion with respect to the wall is controlled by $m(\ell)$, which is directly correlated with $H_{1}(x,\ell)$, the sheet's deformation at leading order (\cref{fig:soft}$c$).
In particular, we find $H_{1}(1;\ell)=(\ell^3-4\ell^2+1.9)/24$, corresponding to a critical motor size $\ell_c \simeq 0.77$ with the foil attracted to the wall for $\ell < \ell_c$ and repelled for $\ell>\ell_c$.
\revision{We note that} attraction or repulsion takes place even though the forces acting on the sheet cancel \rrevision{when integrated in space and averaged in time.}
In fact, it is the deformations at $\OO{\gamma^2}$ in \eqref{eq:soft2} that break the time-reversible symmetry:
\revision{non-time-reversible kinematics is crucial to circumvent the scallop theorem and to generate net motion in viscous flows \cite{Purcell1977,Wiggins1998}.}
Then the effective friction coefficients associated with moving toward and away from the wall are not equal, which works, similar to a ratchet, to enable net average motion.

\begin{figure}
    \includegraphics[width=\linewidth]{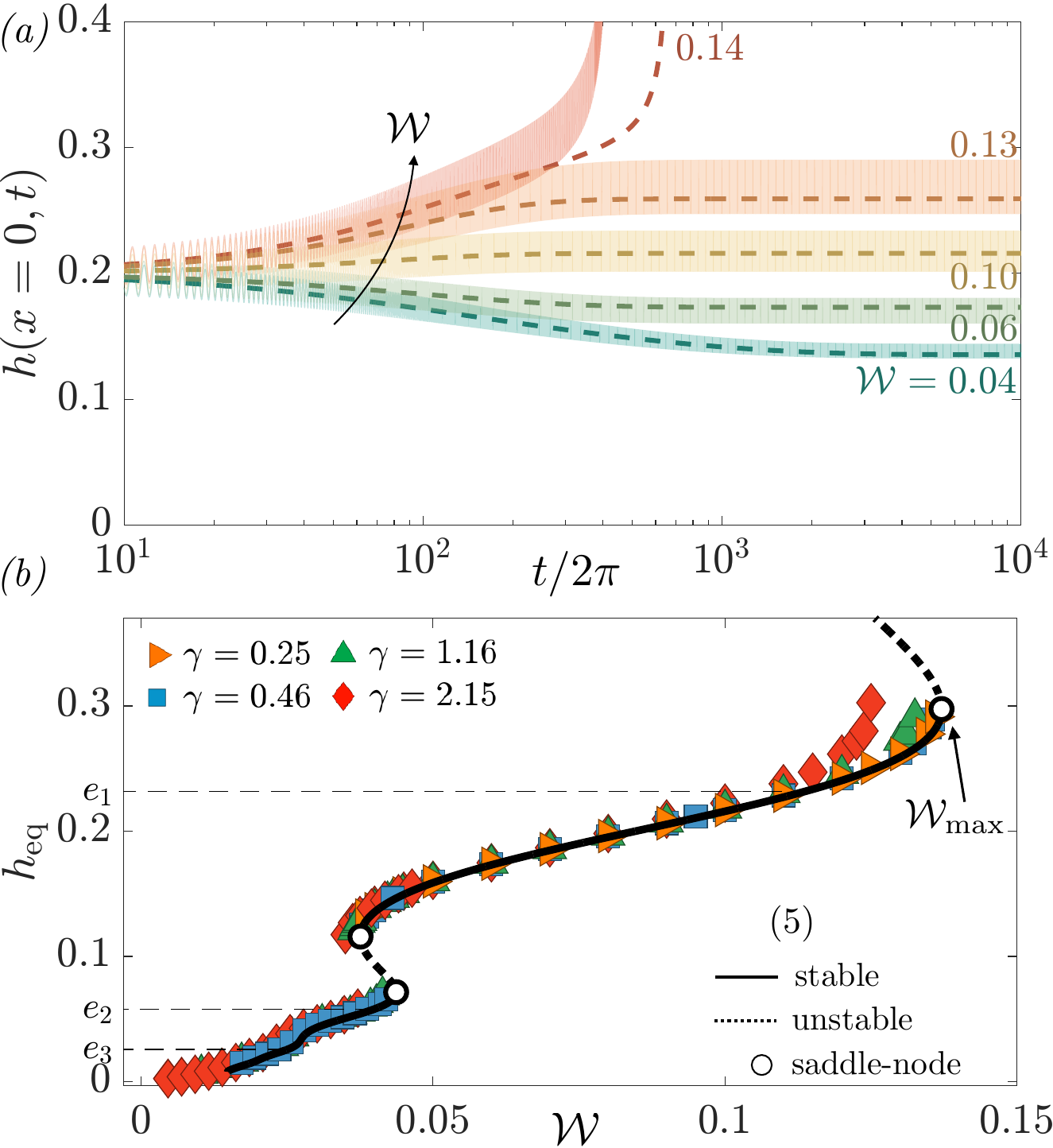}
    \caption{ 
    \revision{$(a)$ Time-evolution of $h(x=0,t)$ for $\gamma=\Gamma=1$. Dashed lines are the numerical solution of \eqref{eq:equili}, shaded lines are the numerical solutions of \eqref{eq:NS} and \eqref{eq:forcebalance}, with the apparent line thickness coming from the sheet's oscillations.}f
    $(b)$ Comparison between the bifurcation diagram obtained by numerical continuation of \eqref{eq:equili} (truncating the sum after $N=5$) with numerical results obtained by solving \eqref{eq:NS} and \eqref{eq:forcebalance} with $\ell=0.05$ (symbols).
    }
\label{fig:regime}
\end{figure}

\revision{We now examine \eqref{eq:softall}  when gravity pulls the sheet away from the wall, $\mathcal W>0$, while the elastohydrodynamic effect acts in the opposite direction for $\ell<\ell_c$.
For heavy sheets with $\mathcal W > 2 m(\ell)$, gravity dominates and the sheet detaches: $f(\tau)\rightarrow 0$, $h(0,t) \rightarrow \infty$.
Conversely, if $\mathcal W < 2 m(\ell)$, attraction dominates and the sheet approaches the wall: $f(\tau) \rightarrow \infty$, $h(0,t) \rightarrow 0$.
}
As such, there is no stable equilibrium height.
However, our numerical simulations reveal a different scenario: as shown in \cref{fig:regime}$(a)$, for small enough weights, the sheet reaches a time-averaged equilibrium height $h_{\rm eq}>0$.
We explain this discrepancy as follows.
\rrevision{The analysis leading to \eqref{eq:softall} assumes $h\sim1$, $\tilde h \sim \tilde H \gg \tilde H_{\rm bv}$. Yet as the sheet moves closer to the wall, eventually $h$ becomes small and  $\tilde h \sim \tilde H_{\rm bv}$. The assumptions behind the previous calculations then break as the viscous and bending stresses become of the same order of magnitude, and a different theoretical approach is required, as discussed next.}

\paragraph{Small distances ($\tilde h \sim \tilde{H}_{\rm bv}$).}
 \rrevision{We now set the heightscale $\tilde H=\tilde H_{\rm bv}$, such that $\Gamma = \gamma$ ($\tilde p_b = \tilde p_v$), with $\mathcal W=\OO{\gamma^0}$}. A direct asymptotic analysis of \eqref{eq:NS} and \eqref{eq:forcebalance} under these assumptions is not feasible, since both the sheet's deformation and the forcing appear at leading order.
Instead, we employ a modal decomposition of the height.
We focus on the limit of an active point force, $\ell\rightarrow 0$, and seek the height as
$h(x,t)=h_0(t)+\gamma\cos(t)H_{1}(x;0)+\gamma^2\mathcal W H_{1}(x;1)+\sum_{i=1}^{N}c_i(t)\zeta_i(x)$,
with $h_0(t)$ and $(c_i(t))_{i=1\dots N}$ to be determined.
The functions $H_{1}(x;0)$ and $H_{1}(x;1)$ are polynomials obtained from the analysis of \eqref{eq:softall}, cf.~\cref{fig:soft}($c$), and describe the leading-order deformations due to an active point force and to a uniform weight, respectively.
The $\zeta_i(x)$ are eigenmodes of the triharmonic operator $\partial^6/\partial x^6$,
which appears when linearizing \eqref{eq:NS} and \eqref{eq:forcebalance} for small deformations (see \cite{supp} and \cref{fig:appendix_mode} for details).
Using this ansatz, we project \eqref{eq:NS} and \eqref{eq:forcebalance} in space and perform a two-timescale asymptotic expansion with the slow time $T=\gamma^2 t$, so that $h_0(t)$ becomes $h_0(t, T)$.
After some algebra \cite{supp}, this yields a differential equation
governing the time-averaged height $\langle h_0 \rangle(T) = \int_t^{t+2\pi}h_0(t,T)~{\rm d}t/2\pi$ at $\OO{\gamma^2}$:
\begin{align}
\begin{split}
	\frac{1}{\langle h_0\rangle ^2} \frac{{\rm d}\langle h_0\rangle }{{\rm d}T}&= 
	\frac{1}{4}\mathcal W \langle h_0\rangle
	- d_0
	+ \sum_{i,j=1}^N d_{ij} g_{ij}\left(\langle h_0\rangle\right),
    \\
	g_{ij}(h)&=\frac{1+\left({h}/{\sqrt{e_ie_j}}\right)^6}{(1+\left({h}/{e_i}\right)^6)(1+\left({h}/{e_j}\right)^6)} .
    \label{eq:equili}
\end{split}
\end{align}
The coefficients $e_i$, $d_{ij}$ and $d_0$ are given in \cite{supp}.
\revision{The first two terms on the r.h.s. of \eqref{eq:equili} recover the analysis in the limit $\tilde h \gg \tilde H_{\rm bv}$, cf.~\eqref{eq:soft} with $f\simeq \langle h_0 \rangle ^{-2}$.}
The sum captures the effect of the modes $\zeta_i$, \revision{which become significant for $\tilde h \sim \tilde H_{\rm bv}$.}  
The coefficients $d_{ij}$ quantify the strength of this contribution, and $e_i$ corresponds to the height scale below which the $i$-th mode is excited: $g_{ij}(\langle h_0 \rangle) \simeq 0$ for $\langle h_0 \rangle \gg e_i,e_j$.

\Cref{eq:equili} admits both stable and unstable equilibria as the weight $\mathcal W$ is varied.
 Using numerical continuation $\cite{Dhooge2003}$, we obtain the bifurcation diagram showing the steady state hovering height $h_{\rm eq}$.
We also numerically integrated the partial differential equations \eqref{eq:NS} and \eqref{eq:forcebalance} up until the averaged height reaches a steady state or diverges.
\Cref{fig:regime} shows that
both approaches closely agree in predicting the stable equilibria \rrevision{for small values of $\gamma$ and up to $\gamma \simeq 1$}.
When $\mathcal W>\mathcal W_{\rm max}\simeq 0.137$, there is no equilibrium and the sheet detaches from the wall, $\langle h_0 \rangle \rightarrow \infty$.
As $\mathcal W$ decreases, a stable equilibrium is created through a saddle-node bifurcation at $\mathcal W = \mathcal W_{\rm max}$.
As $\mathcal W$ further decreases, the equilibrium height continuously decreases.
Appendix A details an analytical study of \eqref{eq:equili}. In short, as the weight decreases and the sheet gets closer to the wall, higher-order modes are excited and create equilibria near $\langle h_0 \rangle \simeq e_i$.
The first branch of the equilibrium curve shown in \cref{fig:regime}, for $h_{\rm eq}>0.1$, corresponds to the excitation of the first mode $\zeta_1(x)$, while the second branch includes progressively higher-order modes as $\mathcal W \rightarrow 0$, $h_{\rm eq}\rightarrow 0$. 
The excitation of higher-order modes allows the sheet to store significant bending energy while keeping deformation amplitudes small when it approaches the wall, preventing contact \revision{and allowing the creation of equilibria for heights near $e_i \tilde H_{\rm bv}$}.
The supplementary movies S3-S5 \cite{supp} illustrate the dynamics and highlight that higher-order modes are indeed prominent for small hovering heights.

\paragraph{Discussion and conclusions.}
Extending our results to circular, axisymmetric sheets yields similar conclusions. The only differences are the coefficients $e_i$, $d_{ij}$ and $d_0$ appearing in \eqref{eq:equili},
leading to a bifurcation diagram similar to that in \cref{fig:regime} (see \cite{supp} for details). 
In particular, for a circular sheet with a Poisson's ratio $\nu=0.3$, we find
\begin{align}
\begin{split}
	\tilde W_{\rm max}&={0.11}\frac{\tilde F_a^2}{(\tilde \mu \tilde \omega \tilde B^{2})^{1/3}}, \\
	\tilde h_{\rm eq}(\tilde W_{\rm max})&=0.19  \left(\frac{\tilde \mu \tilde \omega}{\tilde B}\right)^{1/3} \tilde L^2,
\end{split}
\label{eq:scalings}
\end{align}
\revision{with $\tilde F_a = \tilde m \tilde r \tilde \omega^2$ the active force and $(\tilde \mu \omega \tilde B^2)^{1/3}$  the force scale where bending and viscous forces balance}.
These results are consistent with the scaling results discussed on page 1 using simple arguments.
Although \eqref{eq:equili} and \eqref{eq:scalings} are based on the assumption of a point load, $\ell\to 0$, the scaling is qualitatively the same for a finite ($\ell > 0$) and stiff ($\mathcal B \to \infty$ for $\lvert \xperp \rvert < \ell$) motor, where the difference only enters in the prefactor (see Appendix B). 
Comparing our asymptotic results with the experiments of \cite{Weston2021} using the parameter values described earlier, we find
\revision{$\tilde W_{\rm max}\simeq 30\si{\newton}$ and $\tilde h_{\rm eq}\simeq 2\si{\milli\meter}$},
\revision{of the same order of magnitude as the} reported values $\tilde W_{\rm max}\simeq 5\si{\newton}$
and $\tilde h_{\rm eq}\simeq 0.8\si{\milli\meter}$.
\revision{We explain the overestimation of \eqref{eq:scalings} by two factors. First, the experiments are not performed in the asymptotic regime $\gamma=\tilde F_a/(\tilde \mu \tilde \omega \tilde B^2)^{1/3}  \ll 1$, and we expect a saturation of $\tilde W_{\rm max}$ as $\gamma$ becomes too large \cite{Poulain2025}.
Second,  the Reynolds number constructed using the equilibrium height is $\mathrm{Re}=\OO{10}$. While lubrication theory is known to yield satisfactory results even for such large values, inertial corrections may be needed for refined estimates \citep{Ishizawa1966,Jones1975}.
Despite these limitations, the described model captures the dominant mechanism underlying the hovering of actively driven foils.}


Our analysis of the dynamic interplay between active forcing, viscous fluid flow, and bending stresses demonstrates how a soft foil is attracted to or repelled from a solid surface, depending on the spatial distribution of the forcing.
This mechanism allows the foil to hover while sustaining a substantial weight, akin to a contactless suction cup.
We anticipate that this hovering principle generalizes to a variety of forcing modalities, including active torques, and applies to a wide range of foil sizes and weight-bearing capacities in both air and water, with potential relevance for adhesive behavior in marine organisms ~\cite{chan2019physics}. 
More generally, our findings provide new physical insights into active elastohydrodynamic phenomena and open new avenues for the design of contactless grippers, soft robots, and related technological applications.

\bigbreak
\paragraph{Acknowledgements.}
We thank Sami Al-Izzi, Annette Cazaubiel and Jingbang Liu for insightful discussions.
 S.P and A.C acknowledge funding from the Research Council of Norway through project 341989.
 T.K. acknowledges funding from the European Union's Horizon 2020 research and innovation programme under the Marie Skłodowska-Curie grant agreement No 801133.
 L.M. acknowledges funding from the Simons Foundation and the Henri Seydoux Fund.

\bibliography{biblio}

\appendix

\clearpage

\renewcommand\thefigure{A\arabic{figure}}
\setcounter{figure}{0}

\paragraph{Appendix A: Analysis of \cref{eq:equili} --}

The first three even modes $\zeta_i(x)$ of the harmonic operator $\partial/\partial x^6$ 
{and subject to  $\partial^2\zeta_i/\partial x^2=\partial^3 \zeta_i /\partial x^3=\partial^4 \zeta_i/\partial x^4=0$ at $x=\pm 1$} are shown in \cref{fig:regime}. Their analytical expressions are given in the Supplementary Materials \cite{supp}.
\begin{figure}[!htb]
\includegraphics[width=0.9\linewidth]{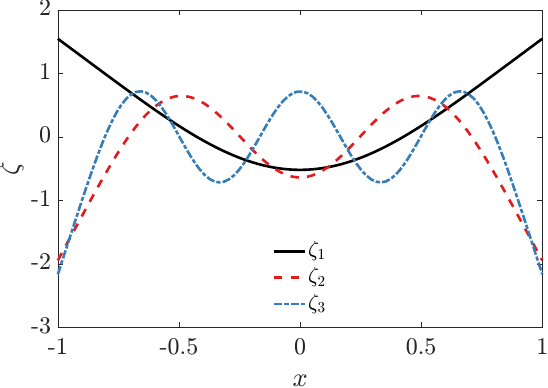}
    \caption{%
    First three modes $\zeta_i(x)$ in the Galerkin projection.}
\label{fig:appendix_mode}
\end{figure}
%

\Cref{eq:equili} without considering the modes $\zeta_i$ (i.e. $N=0$) reads
\begin{align}
    \frac{1}{\langle h_0\rangle ^2} \frac{{\rm d}\langle h_0\rangle }{{\rm d}T}&= 
	\frac{1}{4}\mathcal W \langle h_0\rangle
	- d_0.
    \label{eq:equili_naive}
\end{align}
A linear stability analysis shows that the only fixed point $4d_0/\mathcal W$ is unconditionally unstable.

To analyze \eqref{eq:equili} with the contribution of the modes $\zeta_i$ analytically, we neglect pairwise interactions ($d_{ij}=0$ for $i\neq j$) and assume a scale separation $e_1 \gg e_2 \gg \dots$
We note that $g_{nn}(\langle h_0\rangle)=0$ if $\langle h_0\rangle\gg e_n$, $g_{nn}(\langle h_0\rangle)=1$ if $\langle h_0\rangle\ll e_n$.
Thus, if $\langle h_0 \rangle$ is far from any of the heights $e_n$, the structure of \eqref{eq:equili} is the same as that of \eqref{eq:equili_naive}, and there is no stable equilibrium.
To study the behavior near $e_n$, we write  $\langle h_0\rangle(t)=e_n\left(1+\epsilon_n(t)\right)$, insert in  \eqref{eq:equili}, and expand to \nth{3} order in $\epsilon_n(t)$. We then find that the fixed points of the dynamical system are solutions of the cubic equation:
\begin{align}
	A_n \epsilon_n^3 +B_n \epsilon_n^2 + C_n \epsilon_n+ D_n=0
\label{eq:cubic}
\end{align}
with coefficients
\begin{align*}
\begin{split}
    A_n=4d_{nn}, \quad B_n=\frac{3}{4}d_{nn}, \quad C_n=\frac{\mathcal W}{4}-\frac{3d_{nn}}{2}, \\
    D_n=\frac{\mathcal W}{4}+\frac{d_{nn}}{2}-d_0+\sum_{i=1}^{n-1}d_{ii}.
\end{split}
\end{align*}
The number of solutions of \eqref{eq:cubic} depends on the sign of the discriminant $\Delta_n=18A_nB_nC_nD_n-4B_n^3D_n+B_n^2C_n^2-4A_nC_n^3-27A_n^2D_n^2$: either one solution for $\Delta_n>0$, corresponding to one unstable equilibrium; or three solutions for $\Delta_n<0$, corresponding to two unstable equilibria and one stable equilibrium. The transition between these two behaviors corresponds to saddle-node bifurcations.

\begin{figure}
    \includegraphics[width=\linewidth]{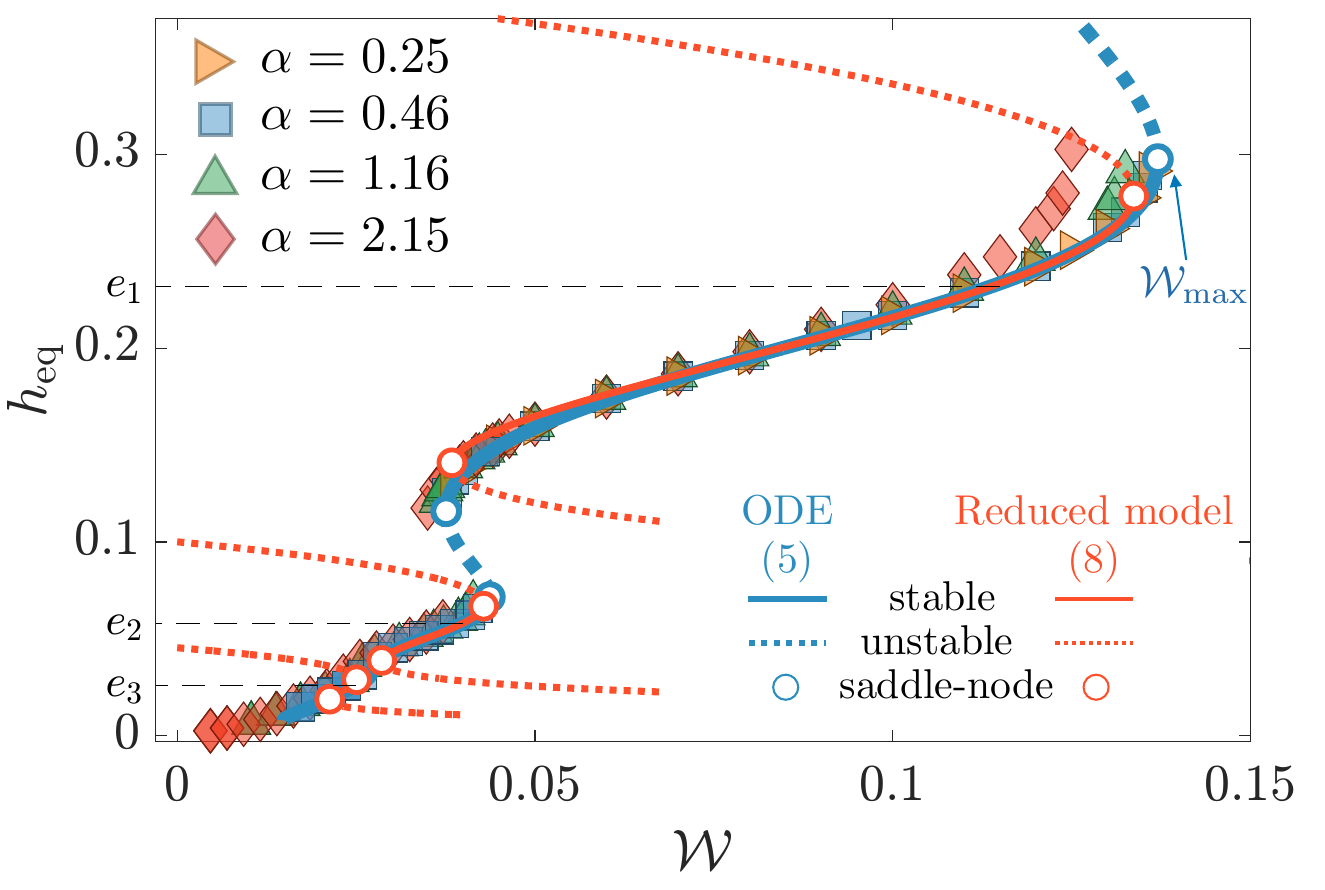}
    \caption{ 
    Comparison between the bifurcation diagram obtained by numerical continuation of \eqref{eq:equili} (blue lines) with numerical results obtained from solving \eqref{eq:NS} and \eqref{eq:forcebalance} with $\ell=0.05$ for different $\gamma$ (symbols).
    }
\label{fig:regimeSupp}
\end{figure}

The bifurcation diagrams predicted from these asymptotic expansions around $e_n$ are shown in \cref{fig:regimeSupp} for $n=1,2,3$, where we note a close agreement with the complete bifurcation diagram of \eqref{eq:soft} obtained numerically.
We observe a cascade of creation and destruction of equilibria as high-order modes get progressively excited for lighter sheets and smaller heights. We note that this is not unlike the snaking bifurcation diagram observed in the Swift-Hohenberg equation \cite{Avitabile2010}, but here arises in a very different setting.
The main discrepancy is that the complete equilibrium diagram only shows two branches. The lower branches, corresponding to $n\geq 2$, are, in fact, all connected. 
This is because the assumption of scale separation is inaccurate for $n\geq 2$ (for example, $e_3/e_2 \simeq 0.44$, $e_4/e_3 \simeq 0.56$). However, the physical picture of higher-order mode excitations as $\mathcal W \rightarrow 0$, $h_{\rm eq} \rightarrow 0$ remains accurate.

\paragraph{Appendix B: Finite motor size --}

\begin{figure}[!htb]
    \includegraphics[width=\linewidth]{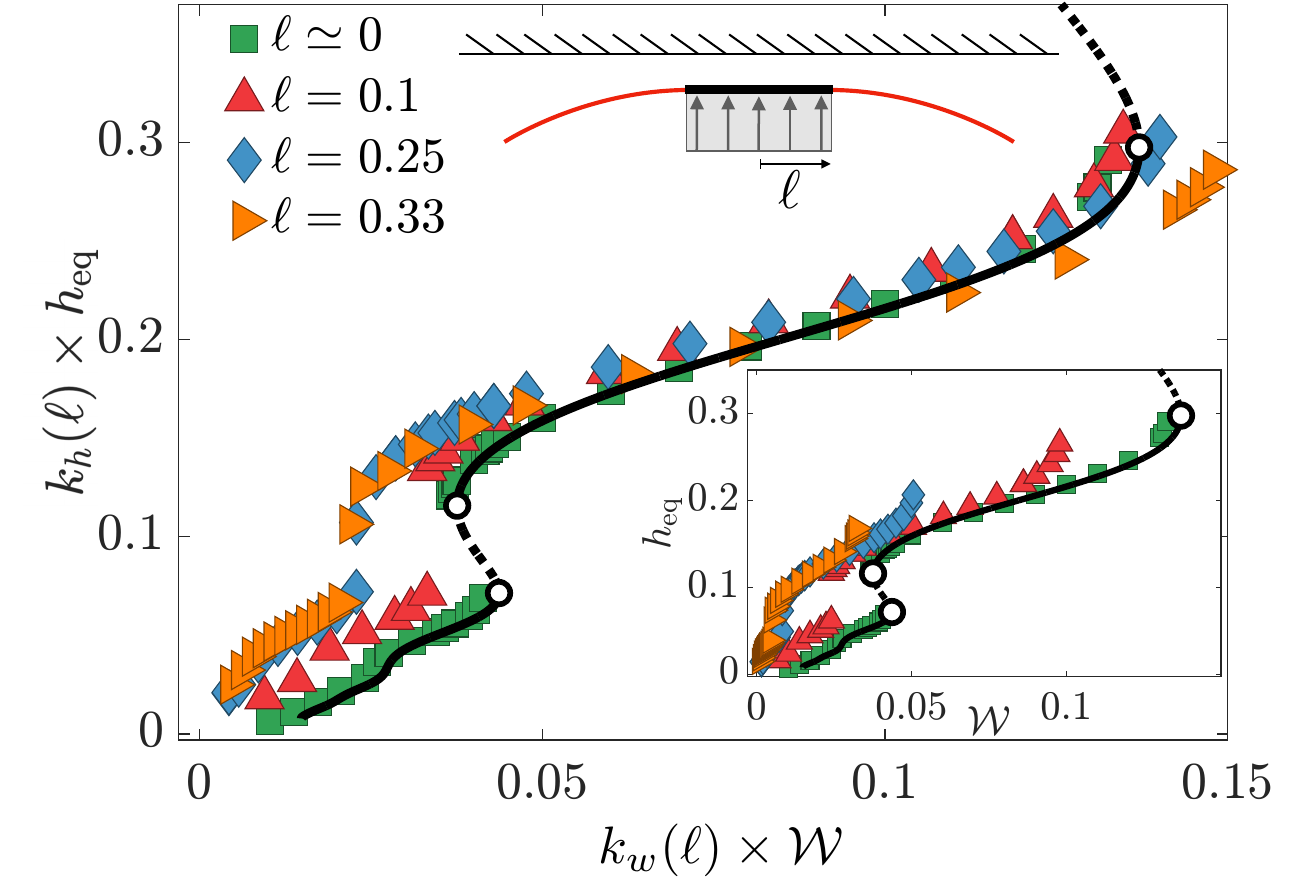}
    \caption{  Equilibrium averaged heights obtained by solving numerically \eqref{eq:NS} and \eqref{eq:forcebalance}  for $\gamma=1.16$ and considering a rigid sheet for $\lvert x \rvert < \ell$. The label $\ell \simeq 0$ corresponds to a uniformly soft sheet with $\ell=0.05$. The results collapse well when rescaling the height and weight to account for the rigid center. The inset shows the data without rescaling.
    Solid lines are the numerical continuation results from solving \eqref{eq:equili}.
    }
\label{fig:motor}
\end{figure}

When an actual motor generates the active forcing, as in the experimental setup of \cite{Weston2021},  it also locally rigidifies the sheet over its area of radius $\tilde \ell$.
We study this effect and solve numerically \eqref{eq:NS} and \eqref{eq:forcebalance} for a sheet rigidified at its center; we obtain the equilibrium curves shown in \cref{fig:motor}. 
The important effect of a finite-size forcing and a locally rigid sheet can be understood using scalings.
From the analysis leading to \eqref{eq:soft}, it is expected that incorporating the finite size effects ($\ell>0$) leads to considering $\gamma^2 m(\ell)$ in place of $\gamma^2$.
Also, a sheet with a rigid domain for $\lvert \tilde x \rvert < \tilde \ell$ is effectively stiffer compared to a soft one, with an effective bending modulus $\tilde B /\left(1-\ell\right)^{4}$.
From these considerations we expect that replacing $h_{\rm eq}$ with $ h_{\rm eq}/k_h(\ell)$ and $\mathcal W$ with $\mathcal W/k_w(\ell)$ collapse the data, with correction factors $k_h(\ell)=(1-\ell)^{4/3}$ and $k_w(\ell)=(m(\ell)/m(0))^2(1-\ell)^{8/3}$. 
\Cref{fig:motor} shows that this is indeed the case. The spread around the second equilibrium curve is expected as the rigid center more strongly affects higher-order modes.
Therefore, for $\ell>0$, \eqref{eq:scalings} becomes
\begin{align}
\begin{split}
	\tilde W_{\rm max}={0.11} \left(\frac{m(\ell}{m(0)}\right)^2\left(1-\ell\right)^{8/3} \frac{\left( \tilde m\tilde r \tilde \omega^2\right)^2}{(\tilde \mu \tilde \omega \tilde B^{2})^{1/3}}, \\
	\tilde h_{\rm eq}(\tilde W_{\rm max})=0.19\left(1-\ell\right)^{4/3} \left(\frac{\tilde \mu \tilde \omega}{\tilde B}\right)^{1/3} \tilde L^2.
\end{split}
\label{eq:scalings_rescaled}
\end{align}

\end{document}


\title{Hovering of an actively driven fluid-lubricated foil -- Supplementary Material}

\author{Stéphane Poulain}
\author{Timo Koch}
\affiliation{%
	Department of Mathematics, University of Oslo, 0851 Oslo, Norway
}
\author{L. Mahadevan}
\affiliation{%
	School of Engineering and Applied Sciences, Harvard University, Cambridge, Massachusetts 02138, USA
}
\affiliation{%
	Departments of Physics, and of Organismic and Evolutionary Biology, Harvard University, Cambridge, Massachusetts 02138, USA
}
\author{Andreas Carlson}
\affiliation{%
	Department of Mathematics, University of Oslo, 0851 Oslo, Norway
}

\maketitle

\renewcommand{\theequation}{S.\arabic{equation}}
\renewcommand{\thefigure}{S.\arabic{figure}}

\section{{Influence of the sheet's tension}}

\rrevision{In Equation (2) of the Main Text we consider stresses associated with the bending of the sheet but neglect those associated with its stretching and that could have two origins.
First, bending deformations that lead to non-zero gaussian curvatures would induce tension within the sheet. 
This effect is absent in 1-dimensional sheets, which bend cylindrically, and is negligible in 2-dimensional sheets when  the deformation amplitude is small compared to the sheet thickness \citep{Landau1986}.
The second possible source of tension comes from the fluid flow. A tangential force balance reads as follows, neglecting the sheet's inertia and written in 1D for simplicity:
$\partial \ddim T/\partial \ddim x = \ddim p \partial \ddim h/\partial \ddim x  + \ddim \mu \partial {\ddim v_x}/\partial{\ddim z}$,  with $\ddim T$ the tension within the sheet and $\tilde v_x$ the horizontal fluid velocity within the thin fluid layer.
The associated contribution of the tension to the normal force balance is $\ddim T \partial^2 \ddim h / \partial \ddim x^2$.
Following the scalings discussed in the main text, $\ddim T \sim \ddim p_v \ddim H$ and $\ddim T \partial^2 \ddim h / \ddim \partial x^2 \sim \ddim p_v \epsilon^2$.
This represents a negligible contribution to the pressure in the lubrication limit $\epsilon^2 \ll 1$ considered herein.}

\section{Boundary conditions}

\label{sec:boundaryConditions}
We consider an elastic sheet with free ends so that its edges have no bending moment, no twisting moment, and no shear force. These three boundary conditions cannot be imposed simultaneously using the Kirchhoff-Love hypotheses. Instead, it is appropriate to use the following (see \cite[p.~586]{Naghdi1973} and references therein for a discussion):
\begin{align}
    \mat M \vect{e_r} \cdot \vect{e_r}= 0, \quad
    \left(\grad_\perp \cdot {\mat M} \cdot\right) \vect{e_r} + \gradperp\left( \mat M \vect{e_r} \cdot \vect{e_\theta}\right)\cdot \vect{e_\theta}=0,
    \label{eq:bc_elastic}
\end{align}
where $\vect{e_r}$ is the outward unit normal at the edge of the sheet and $\vect{e_\theta}$ is the tangent.
In 1D and in 2D axisymmetric cylindrical coordinates (with $r=\lvert \xperp \rvert$), this simplifies to:
\begin{subequations}
\begin{align}
	\textrm{1D:~~at $x=\pm 1$}: &\quad \partialdd hx=0, \quad \partiald{}x\left( \frac\Gamma\gamma\partialdd hx\right)=0,\\
	\textrm{2D~axisymetric~with~$\frac{\Gamma}{\gamma}={\rm cst}:~~$at $r=1$}:& \quad \partialdd hr + \frac \nu r \partiald hr = 0, \quad \partiald{}r\left(\gradperp^2 h\right)=0,
\end{align}
with $\gradperp^2 (.) = \frac1r \partiald{}{r}\left(r\partiald{}{r}(.)\right)$ the harmonic operator in cylindrical coordinates.
In addition, the pressure at the edges recovers the atmospheric pressure so that:
\begin{align}
	\textrm{at $x=\pm 1$ or $r=1$:} \quad p=0.
\end{align}
\label{eq:bcs_force_coordinates}
 \end{subequations}

\section{Numerical method}
We numerically solve (1) and (2) in 1D, written in the form
\begin{subequations}
\begin{align}
	\partiald{}{t}\left(\boldsymbol{S}(\boldsymbol{U})\right)+
	 \partiald{}{x} \left(\boldsymbol{F}(\boldsymbol{U})\right) = \boldsymbol{Q}(\boldsymbol{U}),
\end{align}
 with 
\begin{align}
  \boldsymbol{S} &= \begin{bmatrix} h, & 0, & 0, & 0 \end{bmatrix}, \quad
  \boldsymbol{F} =\begin{bmatrix} 0, &-\dfrac{\partial m}{\partial x}, & -\dfrac{\partial h}{\partial x}, & p \end{bmatrix}, \quad
  \boldsymbol{Q} =\begin{bmatrix} v, & p + f_n(x,t), & {m}\dfrac{\gamma}{\Gamma}, & -\dfrac{12u}{h^2} \end{bmatrix},
\end{align}
and the primary unknowns $\boldsymbol{U}= \begin{bmatrix} v=\partial h/\partial t,& m=({\Gamma}/{\gamma}) \partial^2 h/\partial x^2,& h,& u =q/h,& p  \end{bmatrix}$ representing the vertical sheet velocity, bending moment, height, averaged horizontal fluid velocity, and fluid pressure, respectively. 
The normal load $f_n(x,t)=f_a(x,t)+\gamma \Gamma \mathcal W$ includes the active harmonic forcing and gravity. 
We solve for $x \in [-1,1]$ with boundary conditions
\begin{align}
m = 0, \quad \partiald mx = 0, \quad p&=0.
\end{align}
\end{subequations}
The initial height is constant and systematically varied, while other variables are initially set to zero.
The implementation is realized in the DuMu$^x$ library~\cite{Koch2021}. We discretize in space using a staggered finite volume scheme with pressure unknown at cell centers and other unknowns at vertices; this avoids checkerboard oscillations.
The state variables are advanced in time using a diagonally-implicit Runge-Kutta scheme of order $3$~\cite[Thm.~5]{Alexander1977}, and the nonlinear system at each stage is solved with Newton's method. Lower-order time integration methods either yielded unsatisfactory accuracy or required excessively small time step sizes.
We used $\Delta t = 0.2$ and $0.015\leq\Delta x\leq0.05$ depending on the sheet's softness.
For the results of Fig.~3, simulations ran until the averaged height reached an equilibrium, which could take $t=\OO{10^4}$ up to $t=\OO{10^6}$ depending on the dimensionless parameters.

\section{First-order effect of the sheet's elasticity in 1D}
\label{sec:elasticity}

We consider a 1D uniform soft sheet and neglect gravity.
The governing equations (1) and (2) read:
\begin{subequations}
	\begin{align}
	12\partiald ht-\partiald{}{x}\left(h^3 \partiald px\right)=0, \\
	p=\frac{\Gamma}{\gamma} \partialdddd hx+\frac{\Gamma}{\ell}  \phi_\ell(x) \cos(t) - \gamma \Gamma \mathcal W,	
\end{align}
\label{eq:bendingeasy}
with  $\phi_\ell(x)=1-\heaviside{\lvert x \rvert -\ell}$ and $\mathbb H$ the Heaviside function.
The boundary conditions at $x=\pm 1$ are given by \eqref{eq:bcs_force_coordinates}:
\begin{align}
	\partialdd hx = \partialddd hx = p = 0.
	\label{eq:softbccc}
\end{align}
\label{eq:governingsoft}
\end{subequations}

We use a two-time scale expansion, $t \rightarrow (t,\tau)$, with $t$ the time associated with the forcing and $\tau = \gamma t$ a slow time.
We expand $h$ as
\begin{align}
	h(x,t) =h_0(x,t,\tau)+\gamma h_1(x,t,\tau)+\gamma^2 h_2(x,t,\tau) + \OO{\gamma^3}.
	\label{eq:asymp}
\end{align}
Inserting \eqref{eq:asymp} into \eqref{eq:bendingeasy}, we find at different order in $\gamma$:
\begin{subequations}
\begin{align}
\OO{\gamma^{-1}}:& \quad \partiald{}{x}\left(h_0^3 \partiald{^5 h_0}{x^5}\right)=0,
\label{eq:vibration_orderm1}\\
\OO1:& \quad 12 \partiald{h_0}{t}- \Gamma h_0^3\partialn{6}{h_1}{x} - \frac{\Gamma}{\ell} \cos(t)  h_0^3 \partialdd{\phi_\ell}{x}= 0,
\label{eq:vibration_order0} \\
\OO{\gamma}:& \quad 12 \partiald{h_1}{t} + 12 \partiald{h_0}{\tau} - \Gamma  h_0^3 \partialn{6}{h_2}{x}- 3 \Gamma  h_0^2 \partiald{}{x}\left(h_1\partialn{5}{h_1}{x}\right)- 3 \frac{\Gamma}{\ell}  h_0^2 \cos(t)  \partiald{}{x}\left(h_1\partiald{\phi_\ell}{x}\right)= 0.
\label{eq:vibration_order1}
\end{align}
\end{subequations}
We note that \eqref{eq:vibration_orderm1} gives $h_0=h_0(t,\tau)$,  and we have used this to simplify \eqref{eq:vibration_order0} and \eqref{eq:vibration_order1} already.
The associated boundary conditions, \eqref{eq:softbccc}, at $x=\pm 1$ become:
\begin{subequations}
\begin{align}
\OO1:& \quad \partialdd{h_1}x=\partialddd{h_1}{x}= 0, \quad \partialdddd{h_1}{x}=-\frac{\phi_\ell(1)}{\ell}\cos(t)
\label{eq:bc_order0}, \\
\OO{\gamma}:& \quad \partialdd{h_2}x=\partialddd{h_2}{x}=0, \quad \revision{\partialdddd{h_2}{x}=-{\mathcal W}}.
\label{eq:bc_order1}
\end{align}
\end{subequations}

\subsection{Calculations to $\OO{1}$}
We integrate \eqref{eq:vibration_order0} in space six times, using the boundary conditions \eqref{eq:bc_order0} and the fact that the sheet is symmetric ($\partial h/\partial x=\partial^3h/\partial x^3=\partial^5h/\partial x^5=0$ at $x=0$) to express $h_1$ as:
\begin{subequations}
\begin{align}
    h_1(x,t)&=b(t,\tau)+\cos(t)H_{1,\ell}(x),
    \label{eq:h1softsoft}
\end{align}
where $b(t,\tau)$ is an integration constant to be determined. The function $H_{1,\ell}(x)$ is polynomial by part, given by
\begin{align}
 H_{1,\ell}(x)&=\frac{1}{24}\left(-\frac{3(4\ell-3)x^2}{2}+\frac{(3\ell-2)x^4}{2\ell}-\frac{x^6}{10}+\frac{(x-\ell)^4}{\ell}\heaviside{x-\ell}\right) \quad {\rm for}~x>0.
\end{align}
For $x<0$, the expression is due the parity of $H_{1,\ell}$.
Because the following particular cases of the function $H_{1,\ell}(x)$ will be used later, we write explicitly $H_{1,0}(x)$ corresponding to a point load $\ell=0$, and $H_{1,1}(x)$ corresponding to a uniform load $\ell=1$:
\begin{align}
\begin{split}
	H_{1,0}(x)&=-\frac{x^6}{240}+\frac{x^4}{16}-\frac{x^3}{6}+\frac{3x^2}{16} \quad {\rm for}~x>0, \\
	H_{1,1}(x)&=-\frac{x^6}{240}+\frac{x^4}{48}-\frac{x^2}{16} \quad \forall x.
\end{split}
\label{eq:H10H11}
\end{align}
\end{subequations}

Integrating  \eqref{eq:vibration_order0} and applying the boundary conditions also  gives an ordinary differential equation (ODE) for $h_0$ that admits the solution
\begin{align}
	h_0(t,\tau)&=\left(f(\tau)+\frac{\Gamma}{2}\sin t\right)^{-1/2},
	\label{eq:labeleq}
\end{align}
with $f(\tau)$ an integration constant to be determined.

\subsection{Calculations to $\OO{\gamma}$}
Using \eqref{eq:h1softsoft} and \eqref{eq:labeleq}, \eqref{eq:vibration_order1} becomes:
\begin{align}
	\partialn{6}{h_2}{x}=\left(\frac{12}{\Gamma  h_0^3(t,\tau)}\partiald b t -\frac{6}{\Gamma }f'(\tau)+\frac{9b(t,\tau) \cos(t)}{h_0}\right)+\left(-\frac{12\sin(t)}{\Gamma  h_0^3(t,\tau)}H_{1,\ell}(x)+\frac{9\cos^2(t)}{h_0(t,\tau)}\partiald{}{x}\left(xH_{1,\ell}(x)\right)\right).
\end{align}
We again integrate in space six times, using the boundary conditions \eqref{eq:bc_order1} and symmetry at $x=0$.
This yields an expression for $h_2$ and an ODE for the function $b$ in time $t$:
\begin{align}
    h_2(x,t,\tau)&=\beta(t,\tau) + \frac{\sin (t)}{\Gamma h_0^3(t,\tau)} 12 H_{2,\ell}(x) + \frac{\cos^2(t)}{h_0(t,\tau)} 9 H_{2,\ell}^\star(x) \revision{+\mathcal W H_{1,1}(x)},
    \label{eq:bendingForcing_e1_sol2} \\
    \partiald b t + \frac34\frac{\Gamma  \cos(t)}{f(\tau)+\frac{\Gamma }{2}\sin(t)}b(t,\tau)&=720P_6(\ell)\sin(t) + 630 Q_6(\ell)\frac{\Gamma  \cos^2(t)}{f(\tau)+\frac{\Gamma }{2}\sin(t)}+
    \frac{f'(\tau) \revision{+ \Gamma \mathcal W/2}}{2\left(f(\tau)+\frac{\Gamma }{2}\sin(t)\right)^{3/2}}.
\label{eq:bendingForcing_e1_solalpha}
\end{align}
The function $\beta(t,\tau)$ is an integration constant, and the functions $H_{2,\ell}(x)$, $H_{2,\ell}^\star(x)$ are piecewise polynomials of order 12 whose coefficients depend on $\ell$; $P_6$ and $Q_6$ are 7th-order polynomials. They are given by:
\begin{align}
\begin{split}
     H_{2,\ell}(x)&=P_2(\ell)x^2+P_4(\ell)x^4+P_6(\ell)x^6+\mathcal H_{2,\ell}(x),  \\
     H_{2,\ell}^\star(x)&=Q_2(\ell)x^2+Q_4(\ell)x^4+Q_6(\ell)x^6+\mathcal H_{2,\ell}^\star(x), \\
    \mathcal H_{2,\ell}(x)&=\frac{1}{10!}\left[\frac{45(4\ell-3)x^8}{\num{4}}+\frac{(3\ell-2)x^{10}}{\num{2}\ell}+\frac{x^{12}}{\num{44}}+\frac{(x-\ell)^{10}}{\ell}\heaviside{x-\ell}\right], \\
    \mathcal H_{2,\ell}^\star(x)&=\frac{1}{10!}\left[-\frac{135(4\ell-3)x^8}{\num{4}}+\frac{5(3\ell-2)x^{10}}{\num{2}\ell}-\frac{7x^{12}}{\num{44}}+5\frac{(x-\ell)^{9}(x+\ell)}{\ell}\heaviside{x-\ell} \right],\\
    P_2(\ell)&=\frac{1}{20\times8!}\left(103-210\ell^2+70\ell^4-30\ell^6+10\ell^7\right),~ Q_2(\ell)=\frac{-1}{20\times8!}\left(241-420\ell^2+60\ell^6-30\ell^7\right), \\
    P_4(\ell)&=\frac{-1}{24\times8!}\left(65-140\ell^2+84\ell^4-56\ell^5+12\ell^6\right),~ Q_4(\ell)=\frac{1}{24\times8!}\left(155-280\ell^2+56\ell^5-24\ell^6\right), \\
    P_6(\ell)&=\frac{1}{10!} \left(110-315\ell^2+210\ell^3-63\ell^4+3\ell^6\right),~Q_6(\ell)=\frac{1}{10!} \left(-275+630\ell^2-210\ell^3+6\ell^6\right).
\end{split}
\end{align}
These expressions are valid for $x>0$; for negative $x$, one can use the parity of the height profile.

Equation \eqref{eq:bendingForcing_e1_solalpha} is a first-order linear ODE in $t$ for $b$ that has the solution
\begin{equation}
\begin{split}
    b(t,\tau)=90\left(1+\frac{\Gamma}{2}\sin(t)\right)^{-3/2}&
    \left[ t \left(\frac{1}{2}\frac{{\rm d}f}{{\rm d}\tau}+\revision{\frac{\Gamma \mathcal W}{4}}\right) + 
    8P_6(\ell)\int_0^t\sin(t')\left(f(\tau)+\frac{\Gamma}{2}\sin(t')\right)^{3/2}{\rm d}t'+\right. \\
    & \qquad \quad \left. +  7Q_6(\ell)\Gamma \int_0^t\cos^2(t')\left(f(\tau)+\frac{\Gamma}{2}\sin(t')\right)^{1/2}{\rm d}t'\right].
\end{split}
\label{eq:alpha}
\end{equation}
We choose $f(\tau)$ to ensure that there are no secular terms such that the asymptotic expansion \eqref{eq:asymp} remains valid: this is achieved by requesting  $b$ to be $2\pi-$periodic in $t$ (otherwise $b$ would be unbounded).
The condition $b(2\pi)=b(0)$, with $b(0)=0$, gives an ODE in $\tau$ for $f$.
We approximate the integrals appearing in \eqref{eq:alpha} using a first-order Taylor expansion for small $\Gamma/2f(\tau)$ to find:
\begin{align}
\begin{split}
	f'(\tau)=\Gamma m(\ell) f(\tau)^{1/2} \revision{-\frac{\Gamma \mathcal W}{2}}, \quad
	 m(\ell)=-90\left(6P_6(\ell)+7Q_6(\ell)\right).
	\label{eq:results_1D_PDE}
\end{split}
\end{align}

\section{Stable hovering height in 1D: equilibria for soft sheets with weight}
\label{sec:eqheight}

\label{sec:eqheight}

Here we consider \eqref{eq:governingsoft} using the  heightscale $\tilde H_{bv}\eqdef \tilde L^2\left(\tilde \mu \tilde \omega/\tilde B\right)$, i.e., we let $\tilde H \equiv \tilde H_{\rm bv}$ in all dimensionless expressions involving $\tilde H$. This leads to $\Gamma=\gamma$. 
We also consider $\ell \rightarrow 0$ with $\mathcal W =\OO{\gamma^0}$.
With these considerations, we can write \eqref{eq:governingsoft} as:
\begin{subequations}
\begin{align}
	12\partiald ht-\partiald{}{x}\left(h^3 \partiald px\right)=0,
    \label{eq:alphaa1}\\
	p= \partialdddd hx+\gamma  \phi_0(x) \cos(t) - \gamma^2 \mathcal W,
    \label{eq:alphaa2}\\
     {\rm at}~x=\pm 1: \quad \partialdd hx = \partialddd hx = p = 0.
     \label{eq:alphaa3}
\end{align}
\label{eq:alphaa}
\end{subequations}

\Cref{eq:alphaa1,eq:alphaa2}  without load ($\gamma=0 $) simplify to $\partial h/\partial t -\partial/\partial x \left(h^3\partial^5h/\partial x^5\right)=0$. Linearizing with $h=1+u$, $\lvert u \rvert \ll 1$ gives $\partial u/\partial t - \partial^6 u/\partial x^6=0$. 
This motivates to project \eqref{eq:alphaa}  in space using the eigenfunctions of the tri-harmonic operator subject to appropriate boundary conditions:
\begin{equation}
\begin{split}	
	\partialn{6}{\zeta}{x} + k^6 \zeta = 0 &\quad{\rm for}~-1\leq x\leq1,\\
	\partialdd \zeta x=\partialddd \zeta x=\partialdddd \zeta x=0&\quad{\rm at}~x=\pm 1.
\end{split}
\label{eq:eigenproblem}
\end{equation}
We only seek the even modes. 
The nontrivial solutions $\zeta_n$ to \eqref{eq:eigenproblem} are found by solving an eigenvalue problem. They only exist for discrete values $k_n$ of $k$. For $n$ a positive integer, we find that the associated eigenfunctions $\zeta_n$ are given by:
\begin{align}
\begin{split}
k_n&=n\pi, \\
	\zeta_n(x)&=I_n\left[A_n\cos(k_n x)+B_n\cosh\left(k_n \frac{\sqrt{3}}{2}x\right)\cos\left(\frac{k_n}2x\right)+C_n \sinh\left(k_n \frac{\sqrt{3}}{2}x\right)\sin\left(\frac{k_n}2x\right)\right], \\
		{\rm if}~n~{\rm even}&,
	\begin{cases}
	 A_n&=\frac12 (-1)^{n/2}\cosh\left(\frac{\sqrt 3}{2}k_n\right) \\
	 B_n&=1 \\
	 C_n&=0
	\end{cases};
	\quad {\rm else},
	\begin{cases}
	 A_n&=-\frac12 (-1)^{(n-1)/2}\sinh\left(\frac{\sqrt 3}{2}k_n\right) \\
	 B_n&=0 \\
	 C_n&=1
	\end{cases},
	\label{eq:eigen}
\end{split}
\end{align}
with $I_n$ a coefficient ensuring $\int_{-1}^1 \zeta_n^2(x)~{\rm d}x=1$.
We then seek the solution of \eqref{eq:alphaa} using the following ansatz, with $H_{1,0}$ and $H_{1,1}$ defined earlier in \eqref{eq:H10H11}:
\begin{subequations}
\begin{align}
\begin{split}
	p(x,t)&=\frac32\left(\mathcal W\gamma^2+\gamma  \cos(t)\right)\left(1-x^2\right)+ \gamma  \sum_{i=1}^{N} a_i(t) \zeta_i^{(4)}(x), \\
	h(x,t)&=h_0(t) + \gamma^2 \mathcal W H_{1,1}(x) + \gamma  \cos(t) H_{1,0}(x)+\gamma   \sum_{i=1}^N a_i(t)\zeta_i(x).
\end{split}
\end{align}
These expressions ensure that the normal force balance \eqref{eq:alphaa2} as well as the boundary conditions \eqref{eq:alphaa3} are satisfied, and they recover the analysis of \S  \ref{sec:elasticity} at $\OO{\gamma}$.
The height $h_0(t)$ and the $N$ coefficients $a_i(t)$ are found by projecting \eqref{eq:alphaa1}  on the $\zeta_i$, giving a coupled set of $N+1$ ODEs in $t$:
\begin{align}
	{\rm for}~0\leq i \leq N: \quad \int_0^{1} \left[ 12\partiald ht-\partiald{}{x}\left(h^3\partiald px\right)\right]\zeta_i(x)~{\rm d}x=0,
	\label{eq:integraleq}
\end{align}
\end{subequations}
where we let $\zeta_0(x)=\sqrt{2}/2$. 

\subsection{Projection onto a single mode}

We first focus on describing the dynamics using the first mode $\zeta_1(x)$ only, i.e., $N=1$.
Previous calculations from \S \ref{sec:elasticity} show that the system is more naturally studied after a change of variable: we introduce $F(t)$ as
\begin{align}
	h_0(t)=\left[F(t)+\frac{\gamma}{2}\sin(t)\right]^{-1/2}.
\end{align}
After computing the integrals resulting from $\eqref{eq:integraleq}$ (we perform the averaging in space numerically using Wolfram Mathematica) and expanding in powers of $\gamma$, we obtain differential equations for $F(t)$ and $a_1(t)$.
In particular,
\begin{align}
	\ddt{F}=\gamma  c_0  F^{3/2}(t)\sin(t)+\OO{\gamma^2}, \quad {\rm with}~c_0 \simeq 0.04356
\end{align}
a coefficient found numerically.
This motivates another change of variable: we introduce $\eta(t)$ as
\begin{align}
	\eta(t)=F^{-1/2}(t)-\gamma  \frac{c_0}{2}\cos(t)=\left(h_0^{-2}(t)-\frac{\gamma }2\sin(t)\right)^{-1/2}-c_0\frac{\gamma }{2}\cos(t).
\end{align}
We note that $\eta(t)=h_0(t)+\OO{\gamma}$, and, therefore, $\eta(t)$  directly describes the height evolution.

Using $\eta(t)$ and $a_1(t)$ as the two independent variables, we arrive at the following differential system:
\begin{subequations}
\begin{align}
	\ddt{\eta}&=\gamma^2  \eta^2 \left[-\frac{1}{4}\mathcal W\eta(t) + c_1 + c_2\cos(2t) + c_3 \cos(t)a_1(t) +c_4 a_1(t)^2\right]+\OO{\gamma ^3},
	\label{eq:GalerkinSystemf}\\	\ddt{a_1}&= c_5\sin(t)+c_6\eta^3(t)a_1(t)+\OO{\gamma },	\label{eq:GalerkinSystema}
\end{align}
\end{subequations}
with the $c_i$ constant coefficients found numerically.
Equation \eqref{eq:GalerkinSystemf} shows that ${\rm d}\eta/{\rm d}t=\OO{\gamma^2}$. 
We follow ideas from the averaging method \cite{Sanders2007,Verhulst2023}. We  introduce a slow time $T=\gamma^2 t$, consider $\eta=\eta(T)$ and $a_1=a_1(t,T)$, and apply the averaging operator $\langle u \rangle = \int_{t}^{t+2\pi}u(t')~{\rm d}t'/2\pi$  to \eqref{eq:GalerkinSystemf}. We find:
\begin{align}
	\frac{{\rm d}{\eta}}{{\rm d}T}= \eta^2(T) \left[-\frac{1}{4}\mathcal W \eta(T) + c_1 + c_3 \left\langle \cos(t)a_1(t,T) \right\rangle+c_4 \left\langle a_1(t,T)^2 \right\rangle\right]+\OO{\gamma^3}.
	\label{eq:avgeta}
\end{align}

To describe the height dynamics at $\OO{\gamma^2}$, it suffices to express $a_1(t,T)$ at $\OO1$.
At this order, \eqref{eq:GalerkinSystema} is a linear ODE in $t$ (since $\eta(t)$ becomes $\eta(T)$) that is solved as
\begin{align}
	a_1(t,T)=-c_5 \frac{\cos(t)}{1+c_6^2\eta^6(T)}-c_5c_6\frac{\eta^3(T)\sin(t)}{1+c_6^2 \eta^6(T)}.
	\label{eq:a1Expr}
\end{align}
Another term proportional to $\exp\left[c_6 t \eta^3(T)\right]$ appears in the general solution for $a_1$; we cancel it by setting the associated integration constant to zero.
This amounts to considering the appropriate initial condition for $a_1$ that prevents an initial fast transient dynamic, i.e.,  $a_1(t=0)=-c_5/(1+c_6^2\eta^6(0))$.

Inserting \eqref{eq:a1Expr} into \eqref{eq:avgeta} finally gives an autonomous differential equation that describes the dynamics of the averaged height evolution dynamic at $\OO{\gamma^2}$:
\begin{align}
\begin{split}
	\frac{{\rm d}\eta}{{\rm d}T}=\eta^2\left[\frac{1}{4}\mathcal W\eta - d_0 + \frac{d_1}{1+\left(\dfrac{\eta}{e_1}\right)^6}\right], \quad T=\gamma^2 t,\\
	d_0=0.0122 ,\quad
	d_1=0.0113, \quad
	e_1=0.232,
	\label{eq:equi_N1}
\end{split}
\end{align}
with $d_0$, $d_1$ and $e_1$ constants found from the spatial projection.

\subsection{Projection onto several modes}
The procedure described in the previous section can be applied similarly with $N>1$ modes.
We find that the system of $N+1$ differential equations obtained from  \eqref{eq:integraleq} can be rewritten as:
\begin{align*}
	\ddt{\eta}&=\OO{\gamma^2}, \\
	{\rm for}~1\leq i \leq N:\quad 
	\ddt{a_i}&=k_{1,i}\sin(t)+k_{2,i}\eta^3(t)a_1(t)+\OO{\gamma },
\end{align*}
with the $k_{1,i}$ and $k_{2,i}$ numerical coefficients.
This allows us to use the same averaging procedure: we solve explicitly for the $a_i$ at $\OO{1}$, which gives us the averaged dynamics of $\eta$ at $\OO{\gamma^2}$.
After calculations analogous to the ones shown in the previous paragraph, the solution takes the form
\begin{align}
\begin{split}
	\frac{{\rm d}\eta}{{\rm d}T}&=\eta^2 \left[
	\frac{1}{4}\mathcal W \eta 
	- d_0
	+ \sum_{i=1}^N \frac{d_{ii}}{1+\left(\dfrac{\eta}{e_i}\right)^6}
	+ \sum_{i=1}^N \sum_{j=i+1}^N \frac{d_{ij}\left(1+\left(\dfrac{\eta}{\sqrt{e_ie_j}}\right)^6\right)}{\left(1+\left(\dfrac{\eta}{e_i}\right)^6\right)\left(1+\left(\dfrac{\eta}{e_j}\right)^6\right)}
	\right],\quad T=\gamma^2 t.
	\end{split}
	\label{eq:equiModes}
\end{align}
Up to $N=5$, we find $d_0=\num{1.22e-2}$ and
\begin{align}
\vect e=\begin{pmatrix}
	0.23 \\
	0.058\\
	0.026\\
	0.015 \\
	0.0093
\end{pmatrix}, \quad
\mat d=\begin{pmatrix}
\num{1.14e-2} & \num{2.04e-6} & \num{-6.91e-6}  & \num{1.64e-5} & \num{-3.24e-5 }\\
 & \num{6.95e-4} & \num{9.24e-7} & \num{-2.05e-6} & \num{4.01e-6}  \\
 &  & \num{1.49e-4} & \num{7.08e-7} & \num{-1.24e-6}  \\
  &  & & \num{3.00e-5} & \num{6.12e-7}  \\
    &  & & & \num{4.73e-5}  \\
\end{pmatrix},
\end{align}
where $\vect e$ is the vector with coefficient $e_i$ and $\mat d$  the symmetrical matrix with coefficients $d_{ij}$.

\section{Axisymmetric circular sheets}

Here, we adapt the analysis of a 1D soft sheet to study a 2D circular and axisymmetric sheet. 
The methods and reasoning are identical to those in the 1D case.
We present only the final results below, using the same notations as in \S \ref{sec:elasticity} for the first part when we consider weightless sheets, and of \S\ref{sec:eqheight} for the second part considering the weight of the sheet. We only note that $\Gamma \ell$ becomes $\Gamma \ell^2$ when transposing 1D to 2D, since the normal force balance is considered per unit length in the 1D case and per unit area in the 2D case.

\subsection{Soft weightless sheets}
\label{sec:2Dsoft}
Here, we consider $\mathcal W=0$ and keep the heightscale $H$ arbitrary. The spatial variable is now $ r=\lvert \xperp \rvert$. The governing equations (1) and (2) read:
\begin{subequations}
\begin{align}
	12\partiald ht - \frac1r \partiald{}{r}\left(rh^3\partiald pr\right)=0, \\
	p=\frac{\Gamma}{\gamma} \gradperp^4h+\Gamma \phi_\ell(r)\cos(t),
\end{align}
with boundary conditions at $r=1$ from \eqref{eq:bcs_force_coordinates}:
\begin{align}
	p=0, \quad \partialdd hr + \frac \nu r \partiald hr = 0, \quad \partiald{}r\left(\gradperp^2 h\right)=0.
	\label{eq:axibcsoft}	
\end{align}
\end{subequations}

\begin{figure}[!htb]
\includegraphics[width=0.9\linewidth]{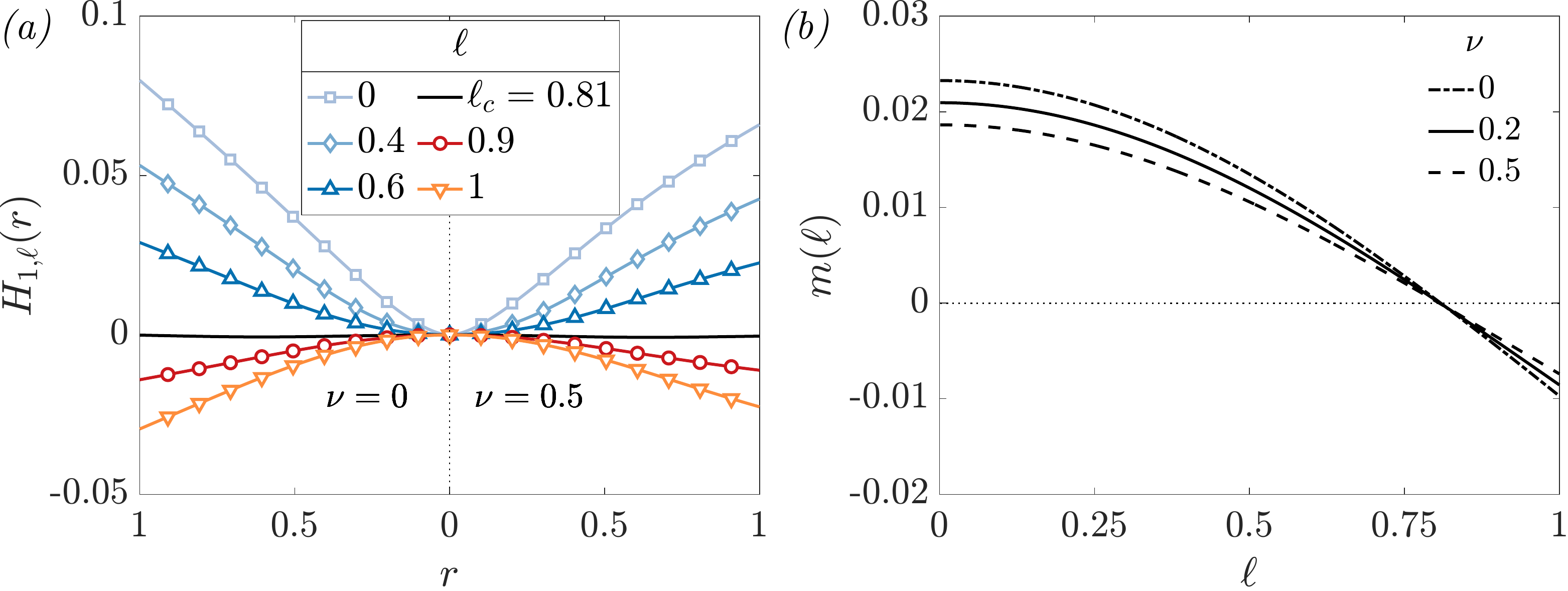};
    \caption{\label{fig:2d}
    Characterization of the dynamics of a circular soft weightless sheet from \eqref{eq:2DSOFT}.
    }
\end{figure}

With $h(x,t)=h_0(x,t,\tau)+\gamma h_1(x,t,\tau)$, $\tau=\gamma t$, we find:
\begin{subequations}
\begin{align}
\begin{split}
	h_0(t)=\left(f(\tau)+\frac{4\Gamma}{3}\sin(t)\right)^{-1/2}, \\
	h_1=\gamma(t,\tau)+\cos(t)H_{1,\ell}(r),
	\end{split}
\end{align}
with 
\begin{align}
\begin{split}
	H_{1,\ell}(r) &= \frac{1}{4}\left(\phi_\nu(\ell) r^2 + \frac{2\ell^2-1}{16\ell^2}r^4 - \frac{r^6}{72} + \frac{\heaviside{r-\ell}}{16}\left[\frac{r^4}{\ell^2}+4r^2-5\ell^2-4\left(\ell^2+2r^2\right)\ln\left(\frac r\ell\right)\right] \right),	 \\
	\phi_\nu(\ell)&=-\frac{(1+5\nu) + 3(1-\nu)\ell^2}{24(1+\nu)}-\frac12 \ln(\ell),\\\end{split}
\end{align}
and
\begin{align}
\begin{split}
	f(\tau)&=\left(1-\frac{m(\ell)}{2} \Gamma  \tau \right)^2, \\
	m(\ell)&=\frac{67+27\nu}{2888(1+\nu)}- \frac{2+\nu}{48(1+\nu)}\ell^2+\frac{\ell^4}{96}-\frac{\ell^6}{576}.
\end{split}
\end{align}
The coefficient $m(\ell)$ and the $\OO{\gamma}$ deformation mode $H_{1,\ell}(r)$ are shown in \cref{fig:2d}. We notice the same correlation discussed in the main text for the 1D case between the sign of $m(\ell)$ and the convexity of $H_{1,\ell}(r)$.
We also note that the Poisson's ratio $\nu$ now appears explicitly in these expressions because of the boundary conditions \eqref{eq:axibcsoft}.
We will specifically need in the next section the deformation for $\ell=0$ and $\ell=1$:
\begin{align}
\begin{split}
		H_{1,1}(r)&=-\frac{r^6}{288}+\frac{r^4}{64}-\frac{2+\nu}{48(1+\nu)}r^2, \\
		H_{1,0}(r)&=-\frac{r^6}{288}+\frac{r^4}{32}+\frac{5+\nu}{96(1+\nu)}r^2-\frac18 r^2\ln(r).
\end{split}
\end{align}
\label{eq:2DSOFT}
\end{subequations}

\subsection{Soft sheets with weight}
\label{sec:2Deq}

Here, we consider $\mathcal W>0$. We follow the same steps as in \S\ref{sec:eqheight}, letting $\tilde H=\tilde H_{bv}$,  $\gamma=\Gamma$. Moreover, $\ell \rightarrow 0$.
We project the height $h(r,t)$ on the eigenfunctions of the tri-harmonic operator in axisymmetric cylindrical coordinates. Factorizing  $(\gradperp^6+k^6)\zeta=(\gradperp^2+k^2)(\gradperp^2-k^2j)(\gradperp^2+k^2j^2)\zeta$, we find that the solutions of $(\gradperp^6+k^6)\zeta=0$  have the form:
\begin{align}
	\zeta_n(r)=A_n J_0(k_nr)+B_n J_0(jk_nr) + C_n J_0(j^2k_nr),
	\label{eq:Bessel}
\end{align}
with $J_m$ the Bessel function of the first kind of $m$-th order and $j=e^{-2i\pi/3}$; the cube roots of $1$ are $1,~j,~j^2$.
The boundary conditions \eqref{eq:axibcsoft} (we use $\gradperp^4 g_n=0$ in lieu of $p=0$) give the following constraints:
\begin{align}
\begin{pmatrix}
J_0(k_n) & -jJ_0(jk_n) & j^2J_0(j^2k_n)\\
J_1(k_n) & -J_1(j k_n) & J_1(j^2 k_n) \\
-kJ_0(k_n)+(1-\nu)J_1(k_n) &
j^2kJ_0(jk_n)-j(1-\nu)J_1(jk_n)&
jkJ_0(j^2k_n)+j^2(1-\nu)J_1(j^2k_n)
\end{pmatrix}
\begin{pmatrix}
A_n \\
B_n \\
C_n
\end{pmatrix}
= 0.
\label{eq:matrixEQ}
\end{align}
The family of $k_n$'s and associated eigenvectors that allow for nontrivial solutions give the corresponding coefficients $A_n~,B_n~,C_n$.
For a given choice of $\nu$, \eqref{eq:matrixEQ} can be solved numerically. For $\nu=0.3$ and 0.5, we find that the first eigenmodes are given by:
\begin{align}
\begin{split}
\nu &=0.3:
\begin{cases}
	k_1=3.8456,
	\quad A_1=1,
	\quad B_1=0.0118 - 0.0793i,
	\quad C_1=\overline B_1, \\
	k_2=7.0022,
	\quad A_2=1,
	\quad B_2=-\num{4.92e-3}-\num{6.68e-4}i,
	\quad C_2=\overline B_2, \\
	k_2=10.1605,
	\quad A_3=1,
	\quad B_3=-\num{4.18e-5}+\num{3.13e-4}i,
	\quad C_3=\overline B_3,
\end{cases} \\
	\nu &=0.5:
	\begin{cases}
	k_1=3.8933,
	\quad A_1=1,
	\quad B_1=0.0107 - 0.0725i,
	\quad C_1=\overline B_1. \\
	k_2=7.0311,
	\quad A_2=1,
	\quad B_2=-\num{4.68e-3}-\num{6.35e-4}i,
	\quad C_2=\overline B_2, \\
	k_3=10.1805
	\quad A_3=1,
	\quad B_3=-\num{4.04e-5}+\num{3.03e-4}i,
	\quad C_3=\overline B_3.
	\end{cases}
\end{split}
\end{align}
The bar denotes the complex conjuguate. We note that the associated eigenfunctions are real-valued.

The ansatz we use for the pressure and height that satisfies the boundary conditions and the normal force balance is then:
\begin{align}
\begin{split}
	p(r,t)&=2\left(\mathcal W\gamma^2+\gamma  \cos(t)\right)\left(1-r^2\right)+\gamma \sum_{i=1}^{N} a_i(t) \gradperp^4g_i(x), \\
		h(r,t)&=h_0(t) + \mathcal W\gamma^2 H_{1,1}(r) + \gamma  \cos(t) H_{1,0}(r)+\gamma \sum_{i=1}^N a_i(t)g_i(r).
\end{split}
\end{align}
The coefficients $h_0$ and $a_i$ are determined by the $(N+1)$ equations: 
\begin{align}
	{\rm for}~0\leq i \leq N: \quad \int_0^{1} \left[ 12\partiald ht-\frac{1}{r}\partiald{}{r} \left(rh^3 \partiald{p}{r}\right)\right]g_i(r)~r{\rm d}r=0,
	\label{eq:integraleqcyl}
\end{align}
where we let $\zeta_0(r)=1$, resulting in a system of $n+1$ ODEs. 

We follow the same averaging and rescaling procedure to arrive at the analog of 	\eqref{eq:equiModes}.
 For 2D sheets, the resulting ODE has the same structure and only differs by the numerical prefactors $(e_i)_{i=1\dots N}$, $d_0$ and $\left(d_{ij}\right)_{i,j=1\dots N}$.
At $N=1$, which suffices to find the maximum supported weight $\mathcal W_{\rm max}$, we find:
\begin{align}
\begin{split}
	\nu &=0.3: \quad
e_1=0.155 
,\quad
d_0 = \num{2.00e-2}, \quad
d_{11}=
\num{1.61e-2}, \\
	\nu &=0.5: \quad
e_1=
	0.151,\quad
d_0 = \num{1.86e-2}, \quad
d_{11}=
\num{1.47e-2}.
\label{eq:coeff2D}
\end{split}
\end{align}

\begin{figure}
\includegraphics[width=0.7\linewidth]{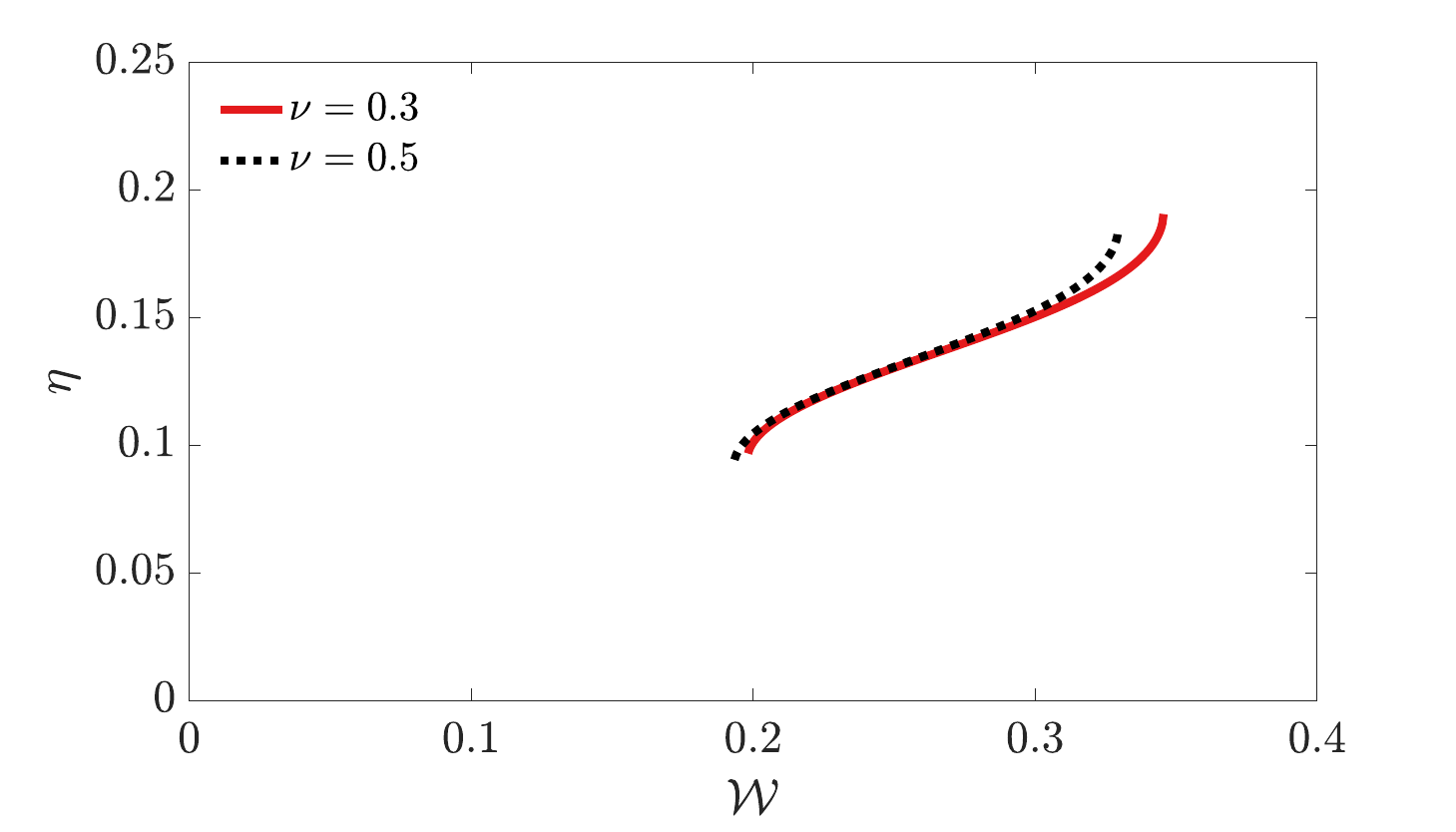}; 
\caption{\label{fig:bifurc2d}
      Stable equilibrium of an axisymmetric obtained from (4) with $N=1$ and coefficients given by \eqref{eq:coeff2D}.
    }
\end{figure}

\section{Movie Descriptions}
{\bf Movies S1 \& S2}: Schematic illustrations of the adhesion (Movie S1, $\ell < \ell_c$) and repulsion (Movie S2, $\ell>\ell_c$) mechanisms in the absence of gravity. The foil's displacement and deformation are computed from the analysis of \cref{sec:elasticity} and amplified for clarity. The pressure profile is computed from the force balance (2) and the velocity vectors are obtained from the height and pressure profiles following classical lubrication theory \cite{Batchelor}.

{\bf Movie S3}: Numerical results ($\gamma=2$, $\ell=0.05$, $\mathcal W=0.13$) in the absence of gravity illustrating how the foil fails to adhere when it is too heavy.

{\bf Movie S4}: Numerical results ($\gamma=2$, $\ell=0.05$, $\mathcal W=0.10$) showing the foil reaching the first equilibrium branch at $h_{\rm eq} \simeq 0.10$.
Once a time-averaged steady state is reached, we analyze the sheet's deformations using a Principal Component Analysis (PCA) on $h'(x,t)=
h(x,t)-\langle h \rangle (x) - \bar h (t) + \langle \bar h \rangle $, where $\langle h \rangle$ is the temporal average and $\bar h$ the spatial average of the height profile. This is defined such that $\langle h' \rangle(x) = \bar{h'}(t) = 0$. 
We show the evolution of $h(x,t)-h'(x,t)$, the combination of the foil's translation mode and of its static deformation due to gravity, and the first spatial modes captured by the PCA of $h'$. In short, the spatial profile of  $h'$ at each timestep is stored in a column of the matrix $\mat H'$. The singular value decomposition $\mat H'=\mat U \mat \Sigma \mat V^\star$ allows to identify the dominant modes corresponding to the largest values of the diagonal matrix $\mat \Sigma$.

{\bf Movie S5}: Numerical results ($\gamma=2$, $\ell=0.05$, $\mathcal W=0.03$ ) showing the foil reaching the second equilibrium branch at $h_{\rm eq} \simeq 0.05$. We perform the same analysis as for Movie S4.

\bibliography{biblio}